\documentstyle[11pt]{article}
        \newcounter{sect}

        \newcommand{\be}{\begin{equation}}
        \newcommand{\ee}{\end{equation}}
        \newcommand{\bea}{\begin{eqnarray}}
        \newcommand{\eea}{\end{eqnarray}}
        \newcommand{\nno}{\nonumber \\}

        \newtheorem{theorem}{Theorem}
        \newtheorem{definition}{Definition}
        \newcommand{\A}{{\cal A}}
        \newcommand{\D}{{\cal D}}\newcommand{\HH}{{\cal H}}
        \newcommand{\LL}{{\cal L}}
        \newcommand{\B}{{\cal B}}
        \newcommand{\K}{{\cal K}}
        \newcommand{\s}{\sigma}
        
        \newcommand{\p}{\partial}
        \newcommand{\dd}{|\D|}
        \newcommand{\n}{\parallel}  
       
        \newcommand{\R}{\mathbf R}
        \newcommand{\C}{\mathbf C}
        \newcommand{\h}{\mathbf H}
\newcommand{\Z}{\mathbf Z}
\newcommand{\N}{\mathbf N}

        \def\slash#1{\rlap{\hskip#1pt\hbox{$\mathchar"236$}}} 
         
        \def\Dslash{{\slash{0.3} D}}

        \def\cross#1{\rlap{\hskip#1pt\hbox{$-$}}}
        \def\intcross{\cross{0.3}\int}
        \def\bigintcross{\cross{2.3}\int}

\begin{document}
        \title{Commutative Geometries are Spin Manifolds}
        \author{A. Rennie\\email: arennie@maths.adelaide.edu.au}
        \maketitle
        \centerline{Abstract}
        In \cite{C1}, Connes presented axioms governing noncommutative 
        geometry. He went on to claim that when specialised to the commutative 
        case, these axioms recover spin or spin$^{c}$ 
        geometry depending on whether the geometry is ``real'' or 
        not. We attempt to flesh out the details of Connes' ideas. 
        As an illustration we present a proof of his claim, partly extending the validity of the result to pseudo-Riemannian 
spin manifolds. Throughout we are as explicit and elementary as possible.
         \section{Introduction}
         \label{intro}
         The usual description of noncommutative geometry takes as its basic data 
         unbounded Fredholm modules, known as $K$-cycles or spectral triples. These 
         are triples $(\A,\HH,\D)$ where $\A$ is an involutive algebra represented on 
         the Hilbert space $\HH$. The operator $\D$ is a closed, unbounded operator 
         on $\HH$ with compact resolvent such that the commutator $[\D,\pi(a)]$ is a 
         bounded operator for every $a\in\A$. Here $\pi$ is the representation of $\A$ in $\HH$. 
         We also suppose that we are given an integer $p$ called the degree of summability 
         which governs the dimension of the geometry. If $p$ is even, the Hilbert space is   
         ${\mathbf Z}_{2}$-graded in such a way that the operator $\D$ is odd.
         \smallskip\newline
         In \cite{C1}, axioms were set down for noncommutative geometry. It is in this framework
         that Connes states his theorem recovering spin manifolds 
         from commutative geometries. Perhaps the most important aspect of this theorem is that 
         it provides sufficient conditions for the spectrum of a $C^{*}$-algebra 
         to be a (spin$^{c}$) manifold. It also gives credence to the idea that spectral triples obeying the
axioms should be regarded as noncommutative manifolds.
\smallskip\newline
Let us briefly describe the central portion of the proof. Showing that the spectrum of $\A$ is 
actually a manifold relies on
the interplay of several abstract structures and the axioms controlling their representation. At the
abstract level we can define the universal differential algebra of $\A$, denoted $\Omega^*(\A)$.
The underlying linear space of $\Omega^*(\A)$ is isomorphic to the chain complex from which we
construct the Hochschild homology of $\A$. In the commutative case, there is also another definition
of the differential forms over $\A$. This algebra, $\widehat{\Omega}^*(\A)$, is skew-commutative
and when the algebra $\A$ is `smooth', \cite{L}, it coincides with the Hochschild homology of $\A$.
For a commutative algebra it is always the case that Hochschild homology contains
$\widehat{\Omega}^*(\A)$ as a direct summand.
\smallskip\newline
The axioms, among other things, ensure that we end up with a faithful representation of
$\widehat{\Omega}^*(\A)$. The process begins by constructing a representation of $\Omega^*(\A)$
from a representation $\pi$ of $\A$, which we may assume is faithful. This is done using the
operator $\D$ introduced above by setting
\be \pi(\delta a)=[\D,\pi(a)]\ \ \ \forall a\in\A,\ee
where $\Omega^*(\A)$ is generated by the symbols $\delta a$, $a\in\A$. There are three
axioms/assumptions controlling this representation. The first is Connes' first order condition. This
demands that $[[\D,\pi(a)],\pi(b)]=0$ for all $a,b\in\A$, at least in the commutative case. It turns
out that the kernel of a representation of $\Omega^*(\A)$ obeying this condition is precisely the
image of the Hochschild boundary. Thus our representation descends to a representation
of Hochschild homology
$HH_*(\A)\subseteq\Omega^*(\A)$, and is moreover faithful. 
\smallskip\newline
The algebra
$\Omega^*_\D(\A):=\pi(\Omega^*(\A))$ is no longer a differential algebra. To remedy this, one
quotients out the `junk' forms, and these turn out to be the submodule 
 generated over $\A$ by graded commutators and
the image of the Hochschild boundary.
Thus the algebra, $\Lambda^*_\D(\A)$, that we arrive at after removing the junk is 
skew-commutative, and we will show
that it is isomorphic to $\widehat{\Omega}^*(\A)$.  We will then prove that the representation of
Hochschild homology with values in $\Lambda^*_\D(\A)$ is still faithful, showing that
$\Lambda^*_\D(\A)\cong\widehat{\Omega}^*(\A)\cong HH_*(\A)$. This is a necessary, 
though not sufficient, condition for the algebra $\A$ to be smooth. Note that by virtue of the first
order condition, both $\Omega^*_\D(\A)$ and $\Lambda^*_\D(\A)$ are symmetric $\A$ bimodules, and so
may be considered to be (left or right) modules over $\A\otimes \A$.
\smallskip\newline
Returning to the axioms, the critical assumption to show that the spectrum is indeed a manifold is
the existence of a Hochschild $p$-cycle which is represented by $1$ or the ${\Z}_2$-grading
depending on whether $p$ is odd or even respectively. The main consequence of this is that this cycle is
nowhere vanishing as a section of $\pi(\Omega^p(\A))$. As it is a cycle, we know that it lies in the
skew-commutative part of the algebra, and this will allow us to find generators of the differential
algebra over $\A$ and construct coordinate charts on the spectrum. Indeed, the non-vanishing of this
cycle is the most stringent axiom, as it enforces the underlying $p$-dimensionality of the spectrum.
\smallskip\newline
Thus we have a two step reduction process
\be \Omega^*(\A)\to\Omega^*_\D(\A)\to\Lambda^*_\D(\A),\ee
and the third axiom referred to above is needed to control the behaviour of the intermediate algebra
$\Omega^*_\D(\A)$, as well as to ensure that our algebra is indeed smooth. Recalling that $\D$ is
required to be closed and self-adjoint, we demand that for all $a\in\A$
\be \delta^n(\pi(a)),\ \ \delta^n([\D,\pi(a)])\mbox{  are bounded for all }n,\ \delta(x)=[\dd,x].\ee
It is easy to imagine that this could be used to formulate smoothness conditions, but it also forces
$\Omega^*_\D(\A)$ to be a (possibly twisted) representation of the complexified Clifford algebra of
the cotangent bundle of the spectrum. As the representation is assumed to be irreducible, we will
have shown that the spectrum is a spin$^c$ manifold. Inclusion of the reality axiom will then
show that the spectrum is actually spin. The detailed description of these matters
requires a great deal of work. 
\smallskip\newline
We begin in Section 2 with some more or less standard background 
results. These will be required in Section 3, where we present our basic definitions and the axioms, as 
well as in Section 4 where we state and prove Connes' result. Section 5 addresses the issue of abstract 
characterisation of algebras having ``geometric representations.'' 

          \section{Background}
          Although there are now several good introductory accounts of noncommutative geometry, eg \cite{La}, 
to make this paper as self-contained as possible, we will quote a number 
          of results necessary for the proof of Connes' result and the analysis of the axioms.
\bigskip

          \subsection{Pointset Topology}

           The point set topology of a compact Hausdorff space $X$ is completely encoded  
           by the $C^*$-algebra of continuous functions on $X$, $C(X)$. This is captured in the 
Gel'fand-Naimark Theorem.

     \begin{theorem}
         For every commutative $C^{*}$-algebra $A$, there exists a Hausdorff 
         space $X$ such that $A\cong C_{0}(X)$. If $A$ is unital, then $X$ is compact.
         \end{theorem}
         In the above, $C_{0}(X)$ means the continuous functions on $X$ which tend to zero at infinity. In
         the compact case this reduces to $C(X)$. We can describe $X$ explicitly as 
$X=Spec(A)=\{$maximal ideals of $A\}=\{$pure states of $A\}$=$\{$unitary equivalence classes of
irreducible representations$\}.$ The weak$^*$ topology on the pure 
state space is what gives us compactness, and translates into the topology of pointwise convergence 
for the states. While this theorem provides much of the motivation for the use of $C^*$-algebras in the context 
of ``noncommutative topology,'' much of their utility comes from the other Gel'fand-Naimark theorem.

         \begin{theorem}
         Every $C^{*}$-algebra admits a faithful and isometric representation 
         as a norm closed self-adjoint 
         $^{*}$-subalgebra of $\B(\HH)$ for some Hilbert space $\HH$.
         \end{theorem}
In the classical (commutative) case there are a number of results showing that we really can 
recover all information about the space $X$ 
from the algebra of continuous functions $C(X)$. For instance, closed sets correspond to norm closed 
ideals, and so single points to maximal ideals. The latter statement is proved using the correspondence 
pure state$\leftrightarrow$kernel of pure state; this is the way that the Gel'fand-Naimark theorem is proved, 
and is not valid in the noncommutative case, \cite{La}. There are many other such correspondences, 
see \cite{WO}, not all of which still make sense in the noncommutative case, for the simple reason
that the three descriptions of the spectrum no longer coincide for a general $C^*$-algebra, \cite{La}.  On this 
cautionary note, let us turn to the most 
important correspondence; the Serre-Swan theorem, \cite{S}.

         \begin{theorem}
         Let $X$ be a compact Hausdorff space. Then a $C(X)$-module $V$ is isomorphic 
         (as a module) to a module $\Gamma(X,E)$ of continuous sections of a complex vector bundle 
         $E\rightarrow X$ if and only if $V$ is finitely generated and projective.
         \end{theorem}
We abbreviate finitely generated and projective to the now common phrase finite projective.  
         This is equivalent to the following. If $V$ is a finite projective $A$-module, then there is 
         an idempotent $e^2=e\in M_{N}(A)$, the $N\times N$ matrix algebra over $A$,  
         for some $N$ such that $V\cong eA^{N}$. Thus $V$ is a 
         direct summand of a free module. We would like then to treat finite projective 
         $C^{*}$-modules as noncommutative generalisations of vector bundles. Ideally we would 
         like the idempotent $e$ to be a projection, i.e. self-adjoint. Since every 
         complex vector bundle admits an Hermitian structure, this is easy to formulate in the 
         commutative case. In the general case we define an Hermitian structure on a right $A$-module $V$ 
         to be a sesquilinear map $\langle\cdot,\cdot\rangle:V\times V\rightarrow A$ such that
         $\forall a,b\in A, v,w\in V$ 
         
         1) $\langle av,bw\rangle=a^*\langle v,w\rangle b$,
         
         2) $\langle v,w\rangle=\langle w,v\rangle^{*}$,
         
         3) $\langle v,v\rangle\geq 0,\ \ \langle v,v\rangle=0\Rightarrow v=0.$
         \smallskip\newline
         A finite projective module always admits Hermitian 
         structures (and connections, for those looking ahead).  Such an Hermitian structure is said to
be nondegenerate if it gives an isomorphism onto the dual module. This corresponds to the usual
notion of nondegeneracy in the classical case, \cite{La}. We also have the following.

         \begin{theorem}
         Let $A$ be a $C^{*}$-algebra. If $V$ is a finite projective $A$-module with a nondegenerate 
         Hermitian structure, then $V\cong eA^{N}$ for some idempotent $e\in M_{N}(A)$ and 
         furthermore $e=e^{*}$, where $^{*}$ is the composition of matrix transposition and the 
         $^{*}$ in $A$.
         \end{theorem}
We shall regard finite projective modules over a $C^*$-algebra as the noncommutative version of 
(the sections of) vector bundles over a noncommutative space.  As a last point before moving on, 
we note that for every 
         bundle on a smooth 
         manifold, there is an essentially unique smooth bundle. For more information on all these results, see 
\cite{La,C}.
\subsection{Algebraic Topology}
Bundle theory leads us quite naturally to $K$-theory in the commutative case, and to further 
demonstrate the utility of defining noncommutative bundles to be finite projective modules, we find 
that this definition allows us to extend $K$-theory to the noncommutative domain as well. Moreover, in 
the commutative case (at least for finite simplicial complexes), $K$-theory recovers the ordinary 
cohomology via the Chern character, and the usual cohomology theories make no sense whatsoever for a noncommutative space. Thus 
$K$-theory becomes the cohomological tool of choice in the noncommutative setting. 
\smallskip\newline
This is not the whole story, however, for we know that in the smooth case we can formulate ordinary cohomology in 
geometric terms via differential forms. 
Noncommutatively speaking, differential forms correspond 
to the Hochschild homology of the algebra of smooth functions on the space, whereas the de Rham cohomology 
is obtained by considering the closely related theory, cyclic homology, \cite{C}. The latter, or more properly 
the periodic version of the theory, see \cite{L,C}, is the proper receptacle for the Chern character 
(in both homology and cohomology).            
\smallskip\newline
We can define $K$-theory for any $C^*$-algebra and $K$-homology for a class of $C^*$-algebras and certain of their dense subalgebras. 
For any of these algebras $A$, elements of $K^*(A)$ may be regarded as equivalence classes of Fredholm modules 
$[(\HH,F,\Gamma)]$. These consist of a representation $\pi:A\rightarrow\B(\HH)$, an operator $F:\HH\rightarrow\HH$ 
such that $F=F^*$, $F^2=1$, and $[F,\pi(a)]$ is compact for all $a\in A$. If $\Gamma=1$, the class defined by the module is 
said to be odd, and it resides in $K^1(A)$. If $\Gamma=\Gamma^*$, $\Gamma^2=1$, $[\Gamma,\pi(a)]=0$ for all $a\in A$ 
and $\Gamma F+F\Gamma=0$, then we call the class even, and it resides in $K^0(A)$. For complex algebras, Bott periodicity 
says that these are essentially the only $K$-homology groups of $A$. For the case of a commutative $C^*$-algebra, $C_0(X)$, 
this coincides with the (analytic) $K$-homology of $X$. A relative group can be defined as well, along with a reduced group 
for dealing with locally compact spaces (non-unital algebras), \cite{BD1,BD2,BDT,BD3}. 
\smallskip\newline
The $K$-theory of the algebra $A$ may be described as equivalence classes of idempotents, $K^0(A)$, and unitaries, $K^1(A)$, in 
$M_\infty(A)$ and $GL_\infty(A)$ respectively. Again, relative and reduced groups can be defined. We will find it useful 
when discussing the cap product to denote elements corresponding to actual idempotents or unitaries 
by $[(e,N)]$, or $[(u,N)]$ respectively, with $e,u\in M_N(A)$. Much of what follows 
could be translated into $KK$-theory or $E$-theory, but we shall be content with the simple presentation below.
\smallskip\newline
The duality pairing between 
$K$-theory and $K$-homology can be broken into two steps. First one uses the cap product, described below, and 
then acts on the resulting $K$-homology class with the natural index map. This map, 
$Index:K^0(A)\rightarrow K^0({\C})={\mathbf Z}$ is given by 
\be Index([(\HH,F,\Gamma)])=Index(\frac{1-\Gamma}{2}F\frac{1+\Gamma}{2}).\ee
On the right we mean the usual index 
of Fredholm operators, and the beauty of the Fredholm module formulation is that the $Index$ map is 
well-defined. The $Index$ map can also be defined on $K^1(A)$, but as the operators in here are all self-adjoint, it
always gives zero. For this reason we will avoid mention of the ``odd'' product rules for the cap 
product.
\smallskip\newline
The cap product, $\cap$, in $K$-theory will allow us to formulate Poincar\'{e} duality in due course. It is a map 
\be \cap:K_*(A)\times K^*(A)\rightarrow K^*(A)\ee
with the even valued terms given on simple elements by the following rules:
\bea & K_0(A)\times K^0(A)\rightarrow K^0(A)\nno
& [(e,N)]\cap [(\HH,F,\Gamma)]=[(e\HH_+^N\oplus e\HH_-^N,\left(\begin{array}{cc} 0 & 
e\tilde{F}^*\otimes 1_Ne\\ e\tilde{F}\otimes 1_Ne & 0\end{array}\right)\left(\begin{array}{cc} e
& 0\\ 0 & -e\end{array}\right))],\eea
where $\HH_\pm=\frac{1\pm \Gamma}{2}\HH$, and $F=\left(\begin{array}{cc}0 & \tilde{F}^*\\ \tilde{F} &
0\end{array}\right)$. 
As every idempotent determines a finite projective module, this is easily seen to be just twisting the 
Fredholm module by the module given by $[(e,N)]$. The 
product of a unitary and an odd Fredholm module will lead to the odd index, 
\bea & K_1(A)\times K^1(A)\rightarrow K^0(A)\nno
 & [(u,N)]\cap [(\HH,F)]=
[((\frac{1+F}{2}\HH)^{2N},\left(\begin{array}{cc} 0 & u\\ u^* & 0\end{array}\right), 
\left(\begin{array}{cc} \frac{1+F}{2}\otimes 1_N & 0\\ 0 & -\frac{1+F}{2}\otimes 1_N\end{array}\right) )].\eea
The cap product turns $K^*(A)$ into a module over $K_*(A)$. 
Usually, one is given a Fredholm module, $\mu=[(\HH,F,\Gamma)]$, where $\Gamma=1$ if the module is odd, 
and then considers the $Index$ as a map on $K_*(A)$ via $Index(k):=Index(k\cap\mu)$, for any $k\in K_*(A)$.  
\smallskip\newline
Let us for a minute suppose that $A$ is commutative.
          If $X=Spec(A)$ is a compact finite simplicial complex, there are isomorphisms
          \be K^{*}(X)\otimes{\mathbf Q}\cong K_{*}(A)\otimes{\mathbf Q}
          \stackrel{ch^*}{\rightarrow} H^{*}(X,{\mathbf Q})\ee
          and
          \be K_{*}(X)\otimes{\mathbf Q}\cong K^{*}(A)\otimes{\mathbf Q}
          \stackrel{ch_*}{\rightarrow} H_{*}(X,{\mathbf Q})\ee
          given by the Chern characters. Here $H^{*}$ and $H_{*}$ are the ordinary 
          (co)homology of $X$. Note that a (co)homology theory for spaces is a homology(co) theory 
for algebras. 
          \smallskip\newline
          Important for us is that these isomorphisms also preserve the cap product, both in $K$-theory 
and ordinary (co)homology, so that the following diagram
          
          \[\begin{array}{lcl}
          K^{*}(X)\otimes{\mathbf Q}
           & \stackrel{\cap\mu}{\rightarrow} 
          & K_{*}(X)\otimes\mathbf{Q} \\
          ch^{*}\downarrow & & ch_{*}\downarrow \\
          H^{*}(X,{\mathbf Q}) & \stackrel{\cap ch_*(\mu)}{\rightarrow} & 
          H_{*}(X,{\mathbf Q})\end{array}\]
commutes for any $\mu\in K_*(X)$. So if $X$ is a finite simplicial complex satisfying 
          Poincar\'{e} Duality in 
          $K$-theory, that is there exists $\mu\in K^*(C(X))$ such that 
$\cap\mu:K_*(C(X))\rightarrow K^*(C(X))$ is an isomorphism, then  
there exists $[X]=ch_{*}(\mu)\in H_{*}(X,{\mathbf Q})$ 
          such that 
          \be \cap[X]:H^{*}(X,{\mathbf Q})\rightarrow H_{*}(X,{\mathbf Q})\ee
          is an isomorphism. If $ch_*(\mu)\in H_p(X,{\mathbf Q})$ for some $p$, then 
we would know that $X$ satisfied Poincar\'{e} duality in ordinary (co)homology, which is 
certainly a necessary condition for $X$ to be a manifold. 
\smallskip\newline
We note in passing that $K$-theory has only even and odd components, whereas the usual (co)homology 
is graded by ${\mathbf Z}$. This is not such a problem if we replace (co)homology with 
periodic cyclic homology(co). In the case of a classical manifold it gives the same results as the usual theory, 
but it is naturally ${\mathbf Z}_2$-graded. Though we do not want to discuss cyclic (co)homology in 
this paper, we note that the appropriate replacement for the commuting square above in the noncommutative case 
is the following.
  
          \[\begin{array}{lcl}
          K_{*}(A)\otimes{\mathbf Q} 
           & \stackrel{\cap\mu}{\rightarrow} 
          & K^{*}(A)\otimes\mathbf{Q} \\
          ch_{*}\downarrow & & ch^{*}\downarrow \\
          H_{*}^{per}(A)\otimes{\mathbf Q} & \stackrel{\cap ch^*(\mu)}{\rightarrow} & 
          H^{*}_{per}(A)\otimes{\mathbf Q}\end{array}\]
Moreover, the periodic theory is the natural receptacle for the Chern character in the 
not necessarily commutative case, provided that $A$ is an algebra over a field containing 
${\mathbf Q}$. For more information on Poincar\'{e} duality in noncommutative geometry, including details of the induced 
maps on the various homology groups, see \cite{C, C3}.

\subsection{Measure}
        On the analytical front we have to relate the noncommutative 
         integral given by the Dixmier trace to the usual measure 
         theoretic tools. This is achieved using two results of 
         Connes; 
         one building on the work of Wodzicki, \cite{W}, and the other on the 
         work of Voiculescu, \cite{V}. For more detailed information on these 
         results, see \cite{C} and \cite{W,V}.
         \smallskip\newline
         To define the Dixmier trace and relate it to Lebesgue 
         measure, we require the definitions of several normed ideals 
         of compact operators on Hilbert space. The first of these is 
         \be 
         \LL^{(1,\infty)}(\HH)=\{T\in\K(\HH):\sum_{n=0}^{N}\mu_{n}(T) 
         =O(\log N)\}\ee
         with norm 
         \be \parallel T\parallel_{1,\infty}=\sup_{N\geq 
         2}\frac{1}{\log N}\sum_{n=0}^{N}\mu_{n}(T).\ee
         In the above the $\mu_{n}(T)$ are the eigenvalues of 
         $|T|=\sqrt{TT^{*}}$ arranged in decreasing order and 
         repeated according to multiplicity so that $\mu_{0}(T)\geq 
         \mu_{1}(T)\geq\ldots$. This ideal will be the domain of 
         definition of the Dixmier trace. Related to this ideal are the ideals 
         $\LL^{(p,\infty)}(\HH)$ for $1<p<\infty$ defined as follows;
         \be 
         \LL^{(p,\infty)}(\HH)=\{T\in\K(\HH):\sum_{n=0}^{N}\mu_{n}(T) 
         =O(N^{1-\frac{1}{p}})\}\ee
         with norm
         \be \parallel T\parallel_{p,\infty}=\sup_{N\geq 
         1}\frac{1}{N^{1-\frac{1}{p}}}\sum_{n=0}^{N}\mu_{n}(T).\ee
         We introduce these ideals because if 
         $T_{i}\in\LL^{(p_{i},\infty)}(\HH)$ for $i=1,\ldots,n$ and 
         $\sum\frac{1}{p_{i}}=1$, then the product $T_{1}\cdots 
         T_{n}\in\LL^{(1,\infty)}(\HH)$. In particular, if 
         $T\in\LL^{(n,\infty)}(\HH)$ then 
         $T^{n}\in\LL^{(1,\infty)}(\HH)$. 
         \smallskip\newline
         We want to define the Dixmier trace so that it returns the coefficient of the 
         logarithmically divergent part of the trace of an operator. 
         Unfortunately, since $(1/\log N)\sum^{N}\mu_{n}(T)$ is in 
         general only a bounded sequence, we can not take the limit in 
         a well-defined way. The Dixmier trace is usually defined in 
         terms of linear functionals on bounded sequences satisfying 
         certain properties. One of these properties is that if the 
         above sequence is convergent, the linear functional returns 
         the limit. In this case, the result is independent of which 
         linear functional is used. So, for 
         $T\in\LL^{(1,\infty)}(\HH)$ with $T\geq 0$, we say that $T$ 
         is measurable if 
         \be \bigintcross
         T:=\lim_{N\rightarrow\infty}\frac{1}{\log N} 
         \sum_{n=0}^{N}\mu_{n}(T)\ee
         exists. 
         Moreover, $\intcross$ is linear on measurable operators, and we 
         extend it by linearity to not necessarily positive operators. 
         Then $\intcross$ satisfies the following properties, \cite{C}:
         
         1)The space of measurable operators is a closed (in the 
         $(1,\infty)$ norm) linear space invariant under conjugation 
         by invertible bounded operators and contains 
         $\LL^{(1,\infty)}_{0}(\HH)$, the closure of the finite rank 
         operators in the $(1,\infty)$ norm;
         
         2) If $T\geq 0$ then $\intcross T\geq 0$;
         
         3) For all $S\in\B(\HH)$ and $T\in\LL^{(1,\infty)}(\HH)$ 
         with $T$ measurable, we have $\intcross TS=\intcross ST$;
         
         4) $\intcross$ depends only on $\HH$ as a topological vector space;
         
         5) $\intcross$ vanishes on $\LL^{(1,\infty)}_{0}(\HH)$.
\smallskip\newline
         In the case 
         that $\A$ is finite dimensional and represented 
         on a finite dimensional Hilbert space, all these ideals of 
         compact operators coincide, and the Dixmier trace reduces to 
         the ordinary trace, \cite{C}. Next we relate this operator theoretic  
definition to geometry.
         \smallskip\newline
  If $P$ is a pseudo differential operator acting on sections 
         of a vector bundle $E\rightarrow M$ over a manifold $M$ of 
         dimension $p$, it has a symbol $\s(P)$. The Wodzicki residue 
         of $P$ is defined by 
         \be 
         WRes(P)= 
         \frac{1}{p(2\pi)^{p}} 
         \int_{S^{*}M}trace_{E}\s_{-p}(P)(x,\xi)\sqrt{g}dxd\xi.\ee
         In the above $S^{*}M$ is the cosphere bundle with respect to 
         some metric $g$, and $\s_{-p}(P)$ is the part of the symbol of 
         $P$ homogenous of order $-p$. In particular, if $P$ is of 
         order strictly less than $-p$, $WRes(P)=0$. The interesting 
         thing about the Wodzicki residue is that although symbols other than 
         principal symbols are coordinate dependent, the Wodzicki residue 
         depends only on the conformal class of the metric \cite{C}.
         It is also a trace on the algebra of 
         pseudodifferential operators, and we have the following 
         result from Connes, \cite{C2,C}.
         \begin{theorem}
         Let $T$ be a pseudodifferential operator of order $-p$ acting 
         on sections of a smooth bundle $E\rightarrow M$ on a $p$ 
         dimensional manifold $M$. Then as an operator on 
         $\HH=L^{2}(M,E)$, $T\in\LL^{(1,\infty)}(\HH)$, $T$ is 
         measurable and $\intcross T=WRes(T)$.
         \end{theorem}
         It can also be shown that the Wodzicki residue is the unique 
         trace on pseudodifferential operators extending the Dixmier 
         trace, \cite{C2}. Hence we can make sense of $\intcross T$ for any 
         pseudodifferential operator on a manifold by using the 
         Wodzicki residue. This is done by setting $\intcross T=WRes(T)$. In particular, 
if $T$ is of order strictly less than $-p=-\dim M$, then $\intcross T=0$. 
This will be important for us later in 
         relation to gravity actions. Before moving on, we note that when we are dealing with the 
noncommutative case there is an extended notion of pseudodifferential operators, symbols 
and Wodzicki residue which reduces to the usual notion in the commutative case; see \cite{CM}.
         \smallskip\newline
         The other connection of the Dixmier trace to our work is its 
         relation to the Lebesgue measure. Since the Dixmier trace 
         acts on operators on Hilbert space we might expect it to be 
         related to measure theory via the spectral theorem. Indeed 
         this is true, but we must backtrack a little into perturbation theory.
\smallskip\newline
      The Kato-Rosenblum theorem, \cite{K}, states that for a self-adjoint 
         operator $T$ on Hilbert space, the absolutely continuous part 
         of $T$ is (up to unitary equivalence) invariant under trace 
         class perturbation. This result does not extend to the joint 
         absolutely continuous spectrum of more than one operator. 
         Voiculescu shows that for a $p$-tuple of commuting self-adjoint 
         operators $(T_{1},\ldots,T_{p})$, the absolutely continuous part of their joint 
         spectrum is (up to unitary equivalence) invariant under 
         perturbation by a $p$-tuple of operators 
         $(A_{1},\ldots,A_{p})$ with $A_{i}\in\LL^{(p,1)}(\HH)$. This 
         ideal is given by 
         \be 
         \LL^{(p,1)}(\HH)=\{T\in\K(\HH): 
         \sum_{n=0}^{\infty}n^{\frac{1}{p}-1}\mu_{n}(T)<\infty\},\ee
         with norm given by the above sum. 
         \smallskip\newline
         Voiculescu, \cite{V}, 
         was lead to investigate, for $X$ a finite subset of 
         $\B(\HH)$ and $J$ a normed ideal of compact operators, the obstruction to 
         finding an approximate unit 
         quasi-central relative to $X$. That is, an approximate unit 
         whose commutators with elements of $X$ all lie in $J$. To do this, 
         he introduced the 
         following measure of this obstruction
         \be k_{J}(X)=\liminf_{A\in R^{+}_{1},A\rightarrow 1}\parallel 
         [A,X]\parallel_{J}.\ee
         Here $R^{+}_{1}$ is the unit interval $0\leq A\leq 1$ in the 
         finite rank operators, and in terms of the norm 
         $\n\cdot\n_{J}$ on $J$, $\parallel[A,X]\parallel_{J} 
         =\sup_{T\in X}\parallel [A,T]\parallel_{J}$. 
         With this tool in hand Voiculescu proves the following 
         result.
              
         \begin{theorem}Let $T_{1},\ldots,T_{p}$ be commuting 
         self-adjoint operators on the Hilbert space $\HH$ and 
         $E_{ac}\subset{\mathbf R}^{p}$ be the absolutely continuous 
         part of their joint spectrum. Then if the multiplicity 
         function $m(x)$ is integrable, we have
         \be 
         \gamma_{p}\int_{E_{ac}}m(x)d^{p}x= 
         (k_{\LL^{(p,1)}}(\{T_{1},\ldots,T_{p}\}))^{p}\ee
         where $\gamma_{p}\in(0,\infty)$ is a constant.
         \end{theorem}
         This result seems a little out of place, 
         as we are using $\LL^{(1,\infty)}$ as our measurable 
         operators. However, Connes proves the following, \cite[pp 
         311-313]{C}.
         
         \begin{theorem}\label{bound}
         Let $D$ be a self-adjoint, invertible, unbounded operator on 
         the Hilbert space $\HH$, and let $p\in(1,\infty)$. Then for 
         any set $X\subset\B(\HH)$ we have
         \be k_{\LL^{(p,1)}}(X)\leq C_{p}(\sup_{T\in X}\parallel 
         [D,T]\parallel)(\bigintcross |D|^{-p})^{1/p},\ee
         where $C_{p}$ is a constant.
         \end{theorem}
The case $p=1$ must be handled separately. In this paper we will be dealing only with 
compact manifolds, and so in dimension $1$, we have only the circle. We will check this 
case explicitly in the body of the proof. So ignoring dimension $1$ for now, we have the following,  
\cite{C}.
         \begin{theorem}
         Let $p$ and $D$ be as above, with 
         $D^{-1}\in\LL^{(p,\infty)}(\HH)$ and suppose that $\A$ is an 
         involutive subalgebra of $\B(\HH)$ such that $[D,a]$ is 
         bounded for all $a\in\A$.Then
         
         1) Setting $\tau(a)=\intcross a|D|^{-p}$ defines a trace on $\A$. 
         This trace is nonzero if $k_{\LL^{(p,1)}}(\A)\neq 0$.
         
         2) Let $p$ be an integer and $a_{1},\ldots,a_{p}\in\A$ commuting 
         self-adjoint elements. Then the absolutely continuous part 
         of their spectral measure
         \be \mu_{ac}(f)=\int_{E_{ac}}f(x)m(x)d^{p}x\ee
         is absolutely continuous with respect to the measure
         \be \tau(f)=\tau(f(a_{1},\ldots,a_{p}))=\bigintcross f|D|^{-p},\ \ 
         \forall f\in C^{\infty}_{c}({\mathbf R}^{p}).\ee
         \label{trace}\end{theorem}
         Combining the results on the Wodzicki residue and these last 
         results of Voiculescu and Connes, we will be able to show 
         that the measure on a commutative geometry is a constant 
         multiple of the measure defined in the usual way. 
\smallskip\newline
         The hypothesis of invertibility used in the above theorems for the 
operator $D$ can be removed provided $\ker D$ is finite dimensional. Then 
we can add to $D$ a finite rank operator in order to obtain an invertible operator, 
and the Dixmier trace will be unchanged. For these purely measure theoretic purposes, 
simply taking $D^{-1}=0$ on $\ker D$ is fine. More care must be taken with $\ker D$ in the 
definition of the associated Fredholm module; see \cite{C}. 
         
 \subsection{Geometry}
         We will avoid discussion of cyclic (co)homology in this paper, as we do 
not absolutely require it. 
More important for this paper,  
is the universal differential algebra construction, and its relation to Hochschild homology.
   \smallskip\newline
         The (reduced) Hochschild homology of an algebra $\A$ with 
         coefficients in a bimodule $M$ is defined in terms of the 
         chain complex $C_{n}(M)=M\otimes\tilde{\A}^{\otimes n}$ with 
         boundary map $b:C_{n}(M)\rightarrow C_{n-1}(M)$
         \bea \lefteqn{b(m\otimes a_{1}\otimes\cdot\cdot\cdot\otimes a_{n})= 
         ma_{1}\otimes a_{2}\otimes\cdot\cdot\cdot\otimes a_{n}}\nno
         &\qquad\qquad &\qquad\quad\ \,+\sum_{i=1}^{n-1}(-1)^{i}m\otimes 
         a_{1}\otimes\cdot\cdot\cdot\otimes a_{i}a_{i+1} 
         \otimes\cdot\cdot\cdot\otimes a_{n}\nno
         &\qquad\qquad &\qquad\quad\  \,+(-1)^{n}a_{n}m\otimes a_{1}\otimes\cdot\cdot\cdot\otimes 
         a_{n-1}\eea
         Here $\tilde{\A}=\A/\mathbf{C}$. In the event that $M=\A$, we 
         denote $C_n(M):=C_n(\A)$ and the resulting homology by $HH_{*}(\A)$, otherwise by 
         $HH_{*}(\A,M)$. Though we will be concerned with the commutative case, the general setting uses 
         $M=\A\otimes\A^{op}$ with bimodule structure $a(b\otimes 
         c^{op})d=abd\otimes c^{op}$, with $\A^{op}$ the opposite algebra 
         of $\A$. With this structure it is clear that 
         \be HH_{*}(\A,\A\otimes\A^{op})\cong \A\otimes\A^{op}\otimes HH_{*}(\A)\cong 
         \A\otimes HH_{*}(\A)\otimes\A^{op}.\label{comm}\ee
We will also require topological Hochschild homology. Suppose we have an algebra $\A$ which is 
endowed with a locally convex and Hausdorff topology such that $\A$ is complete. 
This is equivalent to requiring that for any continuous
semi-norm $p$ on $\A$ there are continuous semi-norms $p',q'$ such that $p(ab)\leq p'(a)q'(b)$ 
for all
$a,b\in\A$. In particular, the product is (separately) continuous. 
Any algebra with a topology given by an infinite family of semi-norms in such a way that 
 the underlying
linear space is a Frechet space satisfies this property, and moreover we may take $p'=q'$ so that
multiplication is jointly continuous. In constructing the topological Hochschild
homology, we use the projective tensor product $\hat{\otimes}$ instead of the usual tensor product,
in order to take account of the topology of $\A$. This is defined by placing on the algebraic tensor
product the strongest locally convex topology such that the bilinear map $(a,b)\to a\otimes b$,
$\A\times\A\to\A\otimes\A$ is continuous, \cite{C,T}. If $\A$ is complete with respect to this 
Frechet topology, then the topological tensor product is also Frechet and complete for this topology.
 The resulting Hochschild homology groups are still denoted by
$HH_*(\A)$. {\em Provided} that these groups are Hausdorff, all the important properties of Hochschild
homology, including the long exact sequence, carry over to the topological setting.
         \bigskip\newline
         The universal differential algebra over an algebra $\A$ is 
         defined as follows. As an $\A$ bimodule, $\Omega^{1}(\A)$ is 
         generated by the symbols $\{\delta a\}_{a\in\A}$ subject 
         only to the relations $\delta(ab)=a\delta(b)+\delta(a)b$. 
         Note that this implies that $\delta({\mathbf C})=\{0\}$. 
         This serves to define both left and right module structures 
         and relates them so that, for instance,
         \be a\delta(b)=\delta(ab)-\delta(a)b.\ee
         We then define 
         \be \Omega^{n}(\A)=\bigotimes^{n}_{i=1}\Omega^{1}(\A)\ee
         and
         \be \Omega^{*}(\A)=\bigoplus_{n=0}\Omega^{n}(\A),\ee
         with $\Omega^{0}(\A)=\A$. Thus $\Omega^{*}(\A)$ is a graded 
         algebra, and we make it a differential algebra by setting 
         \be \delta(a\delta(b_{1})\cdot\cdot\cdot\delta(b_{k}))= 
         \delta(a)\delta(b_{1})\cdot\cdot\cdot\delta(b_{k}),\ee
and
         \be \delta(\omega\rho)=\delta(\omega)\rho+(-1)^{|\omega|}\omega\delta(\rho)\ee
where $\omega$ is homogenous of degree $|\omega|$.
         If $\A$ is an involutive algebra, we make $\Omega^{*}(\A)$ 
         an involutive algebra by setting
         \be (\delta(a))^{*}=-\delta(a^{*}),\ 
         (\omega\rho)^{*}=(\rho)^{*}(\omega)^{*}\ \ a\in\A,\ \rho,\omega\in\Omega^*(\A).\ee
With these conventions, $\Omega^*(\A)$ is a graded differential algebra, with graded 
differential $\delta$.
\smallskip\newline
It turns out, \cite{L}, that the chain complex used to define Hochschild homology $HH_*(\A)$ is the
same linear space as $\Omega^*(\A)$. So
\be C_n(\A)\cong\Omega^n(\A)\ee
\be (a_0,a_1,...,a_n)\to a_0\delta a_1\cdot\cdot\cdot\delta a_n.\ee
The algebraic and differential structures of these spaces are different, however. The relation
between $b$ and $\delta$ is known, and is given by
\be b(\omega\delta a)=(-1)^{|\omega|}[\omega,a]\ee
for $\omega\in\Omega^{|\omega|}(\A)$ and $a\in\A$. We will make much use of this relation in our proof.
\smallskip\newline
For a commutative algebra, there is another definition of differential forms. We retain the definition
of $\Omega^1(\A)$, now regarded as a symmetric bimodule, but define $\widehat{\Omega}^n(\A)$ to be $\Lambda^n_{\A}\Omega^1(\A)$,
the antisymmetric tensor product over $\A$. This has the familiar product 
\be (a_0\delta a_1\wedge\cdot\cdot\cdot\wedge\delta a_k)\wedge(b_0\delta
b_1\cdots\wedge\delta b_m)=a_0b_0\delta a_1\wedge\cdots\wedge\delta a_k\wedge\delta
b_1\wedge\cdots\wedge\delta b_m\ee
and is a differential graded algebra for $\delta$ as above. For smooth (smooth algebras are
automatically 
unital and commutative) algebras, 
see \cite{L} for this technical definition, 
$\widehat{\Omega}^{*}(\A)\cong HH_{*}(\A)$ as graded algebras, \cite{L}, though they have different
differential structures. It can also be shown that the Hochschild homology of any commutative and
unital 
algebra $\A$ contains $\widehat{\Omega}^*(\A)$ as a direct summand, \cite{L}. 
For the smooth functions on a manifold, Connes' exploited the locally convex
topology of the algebra $C^\infty(M)$ and the topological tensor product to prove the analogous theorem for
continuous Hochschild cohomology and de Rham currents on the manifold,\cite{C}.  
\smallskip\newline
We also use the universal differential algebra to define connections in the 
algebraic setting. So suppose that $E$ is a finite projective $A$ module. Then it can be shown, \cite{C}, 
that connections, in the sense of the following definition, always exist. 

\begin{definition}
A (universal) connection on the finite projective $A$ module $E$ is a linear map $\nabla:E\rightarrow \Omega^1(A)\otimes E$ 
such that
\be  \nabla(a\xi)=\delta(a)\otimes\xi+a\nabla(\xi),\ \ \forall a\in A,\ \xi\in E.\ee
\end{definition}
Note that this definition corresponds to what is usually called a universal connection, a connection being 
given by the same definition, but with $\Omega^1(A)$ replaced with a representation of $\Omega^1(A)$
obeying the first order condition. The 
distinction will not bother us, but see \cite{La,C}.  
A connection can be extended to a map $\Omega^*(A)\otimes E\rightarrow\Omega^{*+1}(A)\otimes E$ by 
demanding that $\nabla(\phi\otimes\xi)=\delta(\phi)\otimes\xi+(-1)^{|\phi|}(\phi\otimes 1)\nabla(\xi)$ where $\phi$ is 
homogenous of degree $|\phi|$, and extending by linearity to nonhomogenous terms. 
\smallskip\newline
If there is an Hermitian structure on $E$, and we can always suppose that there is, then we may ask what it means 
for a connection to be compatible with this structure. It turns out that the appropriate condition is 
\be \delta(\xi,\eta)_E=(\xi,\nabla\eta)_E-(\nabla\xi,\eta)_E.\ee
We must explain what we mean here. If we write $\nabla\xi$ as $\sum\omega_i\otimes\xi_i$, then the 
expression $(\nabla\xi,\eta)_E$ means $\sum(\omega_i)^*(\xi_i,\eta)_E$, and similarly for the other term. 
As real forms, that is differentials of 
self-adjoint elements of the algebra, are anti-self adjoint, we see the need for the extra minus sign in the definition. 
Note that some authors have the minus sign on the other term, and this corresponds to their choice of the 
Hermitian structure being conjugate linear in the second variable. 
\smallskip\newline
As a last point while on this subject, if $\nabla$ is a connection on a finite projective $\Omega^*(A)$ module $E$, then 
$[\nabla,\cdot]$ is a connection on $\Omega^*(A)$ ``with values in $E$''. Here the commutator is the $graded$ 
commutator, and our meaning above is made clear by 
\bea[\nabla,\omega]\otimes\xi = \delta\omega\otimes\xi + (-1)^{|\omega|}\omega\nabla\xi
-(-1)^{|\omega|}\omega\nabla\xi\nno
=\delta\omega\otimes\xi. \qquad\qquad\qquad\qquad\qquad\qquad\eea
This will be important later on when discussing the nature of $\D$.
         
         \section{Definitions and Axioms}
         \label{setup}
         We shall only deal with normed, involutive, unital algebras over ${\mathbf C}$ 
         satisfying the $C^*$-condition: $\n aa^*\n=\n a\n^2$. To denote the algebra or ideal
         generated by $a_{1},\ldots,a_{n}$, possibly subject to some relations, 
         we shall write $\langle a_{1},\ldots,a_{n}\rangle$. 
         We write, for $\HH$ a Hilbert space, $\B(\HH),\ \K(\HH)$ respectively 
         for the bounded and compact operators on $\HH$. We write 
$Cliff_{r+s}$, $Cliff_{r,s}$, for the Clifford algebra over ${\R}^{r+s}$ with Euclidean signature, 
respectively signature $(\stackrel{r}{+\cdot\cdot\cdot +}\stackrel{s}{-\cdot\cdot\cdot -})$. 
The complex version is denoted ${\C}liff_{r+s}$. It is important to note that while our algebras $\A$
are imbued with a $C^*$-norm, we do not suppose that it is complete with respect to the topology
determined by this norm, nor that this norm determines the topology of $\A$.
         \smallskip\newline
         %start of definitions 
        Typically in noncommutative geometry, we consider a 
        representation of an involutive algebra
        \be \pi:\A\rightarrow\B(\HH) \ee
        together with a closed unbounded self-adjoint operator $\D$ on $\HH$,
        chosen so that the representation of $\A$ extends to 
        (bounded) representations of $\Omega^{*}(\A)$. This is done by requiring
        \be \pi(\delta a)=[\D,\pi(a)].\ee 
So in particular, commutators of $\D$ with $\A$ must be bounded. We then set
        \be \pi(\delta(a\delta(b_{1})\cdot\cdot\cdot\delta(b_{k}))) = 
        [\D,\pi(a)][\D,\pi(b_{1})]\cdot\cdot\cdot[\D,\pi(b_{k})].\ee
        We say that the representation of $\Omega^{*}(\A)$ is induced 
        from the representation of $\A$ by $\D$. With the 
        $^{*}$-structure on $\Omega^{*}(\A)$ described in the last 
        section, $\pi$ will be a $^{*}$-morphism of $\Omega^{*}(\A)$. We shall frequently 
write $[\D,\cdot]:\Omega^*(\A)\rightarrow\Omega^{*+1}(\A)$, where we mean that the 
commutator acts only on elements of $\A$, not $\delta\A$, as defined above.
       \smallskip\newline
   As $\pi$ is only a $^{*}$-morphism of $\Omega^{*}(\A)$ induced by 
        a $^{*}$-morphism of $\A$, and not a map of differential algebras, 
we should not expect from the 
        outset that it would encode the differential structure of 
        $\Omega^{*}(\A)$. In fact, it is well known that we may have 
        the situation
        \be \pi(\sum_{i} a^{i}\delta b_{1}^{i}\cdot\cdot\cdot\delta 
        b_{k}^{i})=0 \ee
        while
        \be \pi (\sum_{i} \delta a^{i}\delta b_{1}^{i}\cdot\cdot\cdot\delta 
        b_{k}^{i})\neq 0. \ee
        These nonzero forms are known as junk, \cite{C}. To obtain a 
        differential algebra, we must look at 
        \be \pi(\Omega^{*}(\A))/\pi(\delta\ker\pi).\ee
The pejorative term junk is unfortunate, as the example of 
the canonical spectral triple on a spin$^c$ manifold shows (and as we shall show later
 with one extra assumption on $\dd$ and the representation). This 
is given by the algebra of smooth functions $\A=C^{\infty}(M)$ acting as multiplication 
operators on $\HH=L^{2}(M,S)$, where $S$ is the bundle of spinors and $\D$ is taken to be the Dirac operator. 
In this case, \cite{C}, the induced representation of the universal differential algebra is (up to a
possible twisting by a complex line bundle)
        \be \pi(\Omega^{*}(\A))\cong Cliff(T^{*}M)\otimes{\mathbf{C}}={\C}liff(T^*M)\ee
        and
        \be \pi(\Omega^{*}(\A))/\pi(\delta\ker\pi)\cong 
        \Lambda^{*}(T^{*}M)\otimes\mathbf{C}.\ee
        Clearly it is the irreducible representation of the former algebra which 
        encodes the hypothesis `spin$^{c}$'. We will come back to this 
        throughout the paper, and examine it more closely in  Section \ref{pretty}. 
With the above discussion as some kind of 
        motivation, let us now make some definitions.

        \begin{definition} 
        A smooth spectral triple $(\A,\HH,\D)$ is given by a 
        representation 
        \be \pi:\Omega^{*}(\A)\otimes\A^{op}\rightarrow\B(\HH) \ee
        induced from a representation of $\A$ by 
        $\D:\HH\rightarrow\HH$ such that
        
        1) $[\pi(\phi),\pi(b^{op})]=0,\ \forall\phi\in\Omega^*(\A),\ b^{op}\in\A^{op}$
        
        2) $[\D,\pi(a)]\in\B(\HH),\ \forall a\in\A$
        
        3) $\pi(a),[\D,\pi(a)]\in\cap^{\infty}_{m=1}Dom\delta^{m}$
        where $\delta(x)=[\dd,x]$.
        
        Further, we require that $\D$ be a closed self-adjoint operator, such that the 
        resolvent
        $(\D-\lambda)^{-1}$ is compact for all $\lambda\in{\mathbf 
        C}\setminus{\mathbf R}$.
        \end{definition}
The first condition here is Connes' first order condition, and it plays an important r\^{o}le 
in all that follows. It is usually stated as $[[\D,a],b^{op}]=0$, for all $a,b\in\A.$ 
  A unitary change of representation on $\HH$ given by 
        $U:\HH\rightarrow\HH$ sends $\D$ to $U\D 
        U^{*}=\D+U[\D,U^{*}]$. When it is important to distinguish 
        between the various operators so obtained, we will write 
        $\D_{\pi}$. Note that the smoothness condition can be encoded by 
demanding that the map $t\rightarrow e^{it\dd}be^{-it\dd}$ is $C^\infty$ for all 
$b\in\pi(\Omega^*(\A))$. This condition also restricts the possible form of $\dd$ and so $\D$. This
will in turn limit the possibilities for the product struture in $\Omega^*_\D(\A)$, eventually
showing us that it must be the Clifford algebra (up to a possible twisting). 
It is of some importance that, as mentioned, the Hochschild boundary on differential forms is given
by 
\be b(\omega\delta a)=(-1)^{|\omega|}[\omega,a].\ee
A representation of $\Omega^*(\A)$ induced by $\D$ and the first order condition satisfies
\be \pi\circ b=0.\ee
That is, Hochschild boundaries are sent to zero by $\pi$. First this makes sense, as $HH_*(\A)$ is a 
quotient of $C_*(\A)$, which has the same linear structure
as $\Omega^*(\A)$. Second, it gives us a representation of Hochschild homology
\be \pi:HH_*(\A)\to\pi(\Omega^*(\A)).\ee
If $\pi:\A\to\pi(\A)$ is faithful, then the resulting map on Hochschild homology will also be
injective. The reason for this is that the kernel of $\pi$ will be generated solely by elements
satisfying the first order condition; i.e. Hochschild boundaries. We will discuss this matter and its
ramifications further in the body of the proof. 
        \smallskip\newline
        To control the dimension we have two more assumptions. 
        \begin{definition}
        For $p=0,1,2,\ldots$, a $(p,\infty)$-summable spectral triple is a smooth spectral triple 
        with 
        
        1) $\dd^{-1}\in\LL^{(p,\infty)}(\HH)$
        
        2) a Hochschild cycle $c\in 
        Z_{p}(\A,\A\otimes\A^{op})$ with 
        \be \pi(c)=\Gamma \ee
        where if $p$ is odd $\Gamma=1$ and if $p$ is even, 
        $\Gamma=\Gamma^{*}$, $\Gamma^{2}=1$, 
        $\Gamma\pi(a)-\pi(a)\Gamma=0$ for all $a\in\A$ and 
        $\Gamma\D+\D\Gamma=0$.
        \end{definition}
     Note that condition 1) is invariant under unitary change of 
        representation. Condition 2) is a very strict restraint on 
        potential geometries. We will write $\Gamma$ or $\pi(c)$ in all dimensions unless 
we need to distinguish them. As a last definition for now, we define 
        a real spectral triple.
        
         \begin{definition}\label{r}
         A real $(p,\infty)$-summable spectral triple is a 
         $(p,\infty)$-summable spectral triple  
         together with an anti-linear involution $J:\HH\rightarrow\HH$ such that
         
                1)$J\pi(a)^{*}J^{*}=\pi(a)^{op}$
                
                2)$J^{2}=\epsilon,\ J\D=\epsilon'\D J,\ J\Gamma=\epsilon''\Gamma J$,
                 
         where $\epsilon,\epsilon',\epsilon''\in\{-1,1\}$ depend only on $p\bmod 8$ as follows:
                   
         \be\begin{array}{lrrrrrrrr}
         p & 0 & 1 & 2 & 3 & 4 & 5 & 6 & 7 \\
         \epsilon & 1 & 1 & -1 & -1 & -1 & -1 & 1 & 1 \\
         \epsilon' & 1 & -1 & 1 & 1 & 1 & -1 & 1 & 1 \\
         \epsilon'' & 1 & \times & -1 & \times & 1 & \times & -1 & \times 
         \end{array}\label{real}\ee
         \end{definition}
          We will learn much about the involution from the proof, but let us say a few words. First, the map 
$\pi(a)\rightarrow\pi(a)^{op}$ in part $1)$, Definition \ref{r}, is $\C$ linear. Though we won't deal with the noncommutative case 
in any great detail in this paper, let us just point out an interesting feature. The requirement $Jb^*J^*=b^{op}$ 
and the first order condition allow us to say that 
\bea\lefteqn{[\D,\pi(a\otimes b^{op})]=\pi(b^{op})[\D,\pi(a)]+\pi(a)[\D,b^{op}]} \nno
&\qquad &\qquad\ \  =\pi(b^{op})[\D,\pi(a)]+\epsilon'\pi(a)J[\D,\pi(b)]J^*.\eea
This allows us to define a representation $\pi(\Omega^*(A^{op}))=\epsilon'\pi(\Omega^*(A))^{op}$. Then, 
just as $A^{op}$ commutes with $\pi(\Omega^*(A))$, $A$ commutes with $\pi(\Omega^*(A))^{op}$. In the body of
the paper we introduce a slight generalisation of the operator $J$ which allows us to introduce an
indefinite metric on our manifold. We leave the details until the body of the proof. As noted in the
introduction, in the commutative case we find that $\pi(\Omega^*(\A))$ is automatically a symmetric
$\A$-bimodule, so when $\A$ is commutative, we may replace $\A\otimes\A^{op}$ by $\A$.
     \bigskip\newline
         %remaining axioms
         With this formulation (i.e. hiding all the technicalities in the definitions) 
         the axioms for noncommutative geometry are easy to state. A real noncommutative 
         geometry is 
         a real $(p,\infty)$-summable spectral triple satisfying the following two 
         axioms:
  
 \newcounter{ax}
         \addtocounter{ax}{1}
         \roman{ax})  \underline{Axiom of Finiteness and Absolute Continuity}
         \addtocounter{ax}{1}
         \newline
         As an $\A\otimes\A^{op}$ module, or equivalently, as an $\A$-bimodule, 
         $\HH_{\infty}=\cap_{m=1}^{\infty}\mbox{Dom}\D^{m}$ 
         is finitely generated and projective. Writing $\langle\cdot,\cdot\rangle$ 
         for the inner product on $\HH_{\infty}$, we require that there be given an 
Hermitian structure, $(\cdot,\cdot)$, on $\HH_{\infty}$ such that 
         \be\langle a\xi,\eta\rangle=\bigintcross a(\xi,\eta)|\D|^{-p},\ \forall a\in\pi(\A),\ 
         \xi,\eta\in\HH_{\infty}.\ee
         
         \roman{ax})  \underline{Axiom of Poincar\'{e} Duality}
         \setcounter{ax}{1}
         \newline
         Setting $\mu=[(\HH,\D,\pi(c))]\in KR^{p}(\A\otimes\A^{op})$, we require that the cap 
         product by $\mu$ is an isomorphism;
         \be K_{*}(\A)\stackrel{\cap\mu}{\rightarrow} K^{*}(\A).\ee 
         Note that $\mu$ depends only on the homotopy class of $\pi$, and in particular is 
         invariant under unitary change of representation. If we want 
         to discuss submanifolds (i.e. unfaithful representations of 
         $\A$ that satisfy the definitions/axioms) we would have to consider $\mu\in K^{*}(\pi(\A))$, and 
         the isomorphism would be between the $K$-theory and the 
         $K$-homology of $\pi(\A)$. Equivalently, we could consider 
         $\mu\in K^*(\A,\ker\pi)$, the relative homology group, \cite{BDT}. We will not require 
this degree of generality. To see that $\mu$ defines a $KR$ class, note that $J\cdot J^*$ 
provides us with an involution on $\pi(\Omega^*(\A\otimes\A^{op}))$. It may or may not be trivial, but 
always allows us to regard the $K$-cycle obtained from $\mu$ as Real, in the sense of \cite{C3}.
            \bigskip\newline
         %Discussion of axioms.
        Axiom \roman{ax}) controls a great deal of the topological and measure theoretic 
         structure of our geometry. Suppose $\A$ is a commutative algebra. Since 
         $\overline{\pi(\A)}$ is a $C^{*}$-subalgebra of $\B(\HH)$, any finite projective 
         module over  $\overline{\pi(\A)}$ is isomorphic to a bundle of continuous sections 
         $\Gamma(X,E)$ for some complex bundle $E\rightarrow X$. Here $X=Spec(\overline{\pi(\A)})$.
         However, here we have a $\pi(\A)$ module, and this distinction is tied up with the 
         smoothness of the coordinates. In particular, this axiom tells us that the algebra $\A$ is
complete with respect to the topology determined by the semi-norms $a\to\n\delta^n(a)\n$. To see this,
consider the action of the completion, which we temporarily denote by $\A_\infty$, on $\HH_\infty$. If
$\A$ were not complete, then $\A_\infty\HH_\infty\not\subseteq\HH_\infty$, because being finite and
projective over $\A$, $\HH_\infty=e\A^N$, for some idempotent $e\in M_N(\A)$ and some $N$. 
However, $\HH_\infty$
is defined as the intersection of the domains of $\D^m$ for all $m$. In particular, $\D^2$ preserves
$\HH_\infty$, so that $\dd$ must also. If we write $\D=F\dd=\dd F$, where $F$ is the phase of $\D$,
then $F$ too must preserve $\HH_\infty$. So let $a\in\A_\infty$ and $\xi\in\HH_\infty$. Then 
\be D^ma\xi=F^{m\bmod 2}\dd^ma\xi\ee
and by the boundedness of $\delta^m(a)$ for all $m$, we see that $D^ma\xi\in\HH_\infty$ for all $m$.
Hence $\A_\infty=\A$, and $\A$ is complete. We shall continue to use the symbol $\A$, and note that
the completeness of $\A$ in the topology determined by $\delta$ makes $\A$ a Frechet space and allows
us to use the topological version of Hochschild homology.
         \bigskip
         \addtocounter{ax}{1}
         \smallskip\newline
         Axiom \roman{ax}), perhaps surprisingly, is related to the Dixmier trace.
          Connes has shown, \cite{C}, that the Hochschild cohomology class (but importantly, {\em not} 
the cyclic class, see \cite{CM,C4}) of $ch_{*}([(\HH,\D,\pi(c))])$ 
          is given by $\phi_{\omega}$, where
          \be\phi_{\omega}(a_{0},a_{1},\ldots,a_{p})= \lambda_{p}\bigintcross\pi(c)a_{0}[\D,a_{1}]
          \cdots[\D,a_{p}]|\D|^{-p}\ee
          for any $a_{i}\in\pi(\A)$. Here $\pi(c)$ is the representation of the Hochschild cycle and 
          $\lambda_{p}$ is a constant. Since the $K$-theory pairing 
          is non-degenerate by Poincar\'{e} Duality, 
          $\phi_{\omega}\neq 0$. Thus, in particular, 
          $\intcross\pi(c)^{2}|\D|^{-p}=\intcross|\D|^{-p}\neq 0$, and operators 
          of the form \be\pi(c)a_{0}[\D,a_{1}]\cdots[\D,a_{p}]|\D|^{-p}\ee are 
          measurable; i.e. their Dixmier trace is well-defined. This also shows us , for example, 
          that elements of the form $\pi(c)^{2}a|\D|^{-p}=a|\D|^{-p}$ are measurable for all 
          $a\in\pi(\A)$ and so defines a trace on $\pi(\A)$. It also tells us that $\dd^{-1}\notin \LL^{(p,\infty)}_0(\HH)$, and furthermore, 
that the cyclic cohomology class is not in the image of the periodicity operator; i.e. $p$ is a lower bound on the 
dimension of the cyclic cocycle determined by the Chern character of the Fredholm module associated to 
$\mu$, \cite{C}. 
For more details on $K$-theory and Poincar\'{e} duality, see 
          \cite{C,BD1,BD2,BDT,BD3}.
          \smallskip\newline
          Though the representation of $\D$ may vary a great deal within the $K$-homology class 
          $\mu$, the trace defined on $\A$ by $\intcross\cdot|\D|^{-p}$ 
 is invariant under unitary change of representation. Let $U\in\B(\HH)$ be 
          unitary with $[\D,U]$ bounded. Then
          \be 
          |U\D U^{*}|=\sqrt{(U\D U^{*})^{*}(U\D U^{*})}=\sqrt{U\D^{2}U^{*}}.\ee
          However, 
          $(U|\D|U^{*})^{2}=U|\D|^{2}U^{*}=U\D^{2}U^{*}$ so 
          \be |U\D U^{*}|=U|\D|U^{*}.\ee
          From this $(U|\D|U^{*})^{-1}=U|\D|^{-1}U^{*}$, and
          \be|U\D U^{*}|^{-p}=U|\D|^{-p}U^{*}.\ee
          It is a general property of the Dixmier trace, \cite{C}, that conjugation by bounded invertible 
          operators does not alter the integral (this is just the trace property). So 
          \be \bigintcross|U\D U^{*}|^{-p}=\bigintcross 
          U|\D|^{-p}U^{*}=\int|\D|^{-p}.\ee
          Furthermore, for any $a\in\A$,
          \be \bigintcross U\pi(a)U^{*}|U\D U^{*}|^{-p}=\bigintcross 
          \pi(a)|\D|^{-p},\ee
          showing that integration is well-defined given
          
          1) the unitary equivalence class $[\pi]$ of $\pi$
          
          2) the choice of $c\in Z_{p}(\A,\A\otimes\A^{op})$.
\smallskip\newline
     For this reason we do not need to distinguish between 
          $\intcross|\D_{\pi}|^{-p}$ and $\intcross|\D_{U\pi U^{*}}|^{-p}$ 
          and we will simply write $\D$ in this context. 
          Anticipating our later interest, we note that  $\intcross|\D|^{2-p}$
          is not invariant.  Sending $\D$ to $U\D U^*$ sends $\intcross \dd^{2-p}$ to 
\bea 
\lefteqn{\bigintcross (U\D U^*)^2|U\D U^*|^{-p} =  \bigintcross (\D+A)^2U\dd^{-p}U^*}\nno
 & \qquad\qquad &\qquad\qquad  =\bigintcross (\D^2 +\{\D,A\}+A^2)U\dd^{-p}U^*\nno
 & \qquad\qquad &\qquad\qquad =\bigintcross (\D^2U\dd^{-p}U^*)+
\bigintcross(\{\D,A\}+A^2)\dd^{-p}\nno
& \qquad\qquad & \qquad\qquad =\bigintcross(\D^2+\{\D,A\}+A^2)\dd^{-p}-\bigintcross U[\D^2,U^*]\dd^{-p},\eea
where $A=U[\D,U^*]$ and $\{\D,A\}=\D A+A\D$.  For this to make sense we must have $[\D,U^*]$
bounded of course. 
         It is important for us that we can evaluate this using the 
          Wodzicki residue when $\D$ is an operator of order $1$ on a 
          manifold. Note that when this is the case, $U[\D^2,U^*]$ is a first order operator, and the
contribution from this term will be from the zero-th order part of a first order operator.
          
           \setcounter{ax}{1}

\newpage
          %theorem and proof   
         \section{Statement and Proof of the Theorem}
         \label{guts}
\subsection{Statement}
         \begin{theorem}[Connes, 1996]
              Let $(\A,\HH,\D,c)$ be a real, $(p,\infty)$-summable 
              noncommutative geometry with $p\geq 1$ such that
              
              \roman{ax}) $\A$ is commutative and unital;
              \addtocounter{ax}{1}
              
              \roman{ax}) $\pi$ is irreducible (i.e. only scalars commute with $\pi(\A)$ 
              and $\D$).
              
          Then
          
              1) The space $X=Spec(\overline{\pi(\A)})$ is a compact, 
connected, metrisable  Hausdorff space for the weak$^*$ topology. So $\A$ is
separable and in fact finitely generated.
              
              2) Any such $\pi$ defines a metric $d_{\pi}$ on $X$ by
              \be d_{\pi}(\phi,\psi)=\sup_{a\in \A}\{|\phi(a)-\psi(a)|:
              \|[\D,\pi(a)]\|\leq 1\}\label{met}\ee
and the topology defined by the metric agrees with the weak$^*$ topology. 
              Furthermore this metric depends only on the unitary equivalence class of $\pi$.
              
              3) The space $X$ is a smooth spin manifold, and the metric above agrees with that 
              defined using geodesics.  For any such $\pi$ there is a smooth embedding 
$X\hookrightarrow{\mathbf R}^{N}$.

              4) The fibres of the map $[\pi]\rightarrow d_{\pi}$ are a finite 
                 collection of affine spaces $A_{\sigma}$ parametrised by the spin 
                 structures $\sigma$ on $X$.
                
              5) For $p>2$, $\intcross|\D_{\pi}|^{2-p}:=WRes(|\D_{\pi}|^{2-p})$ is a 
              positive quadratic 
              form on each $A_{\sigma}$, with unique minimum $\pi_{\sigma}$.
              
              6) The representation $\pi_{\sigma}$ is given by $\A$ acting as 
              multiplication operators on the Hilbert space $L^{2}(X,S_{\sigma})$ and $\D_{\pi_\sigma}$
 as the 
              Dirac operator of the lift of the Levi-Civita connection 
              to the spin bundle 
              $S_{\sigma}$.
              
              7)  For $p>2$ $\intcross|\D_{\pi_{\s}}|^{2-p}=-\frac{(p-2)c(p)}{12} \int_{X}R\sqrt{g}d^{p}x$ where 
              $R$ is the scalar curvature  and 
              \be c(p)=\frac{2^{[p/2]}}{(4\pi)^{p/2}\Gamma(p/2+1)}\label{const}.\ee
              
          \end{theorem}
          %end of statement
          {\it Remark}: Since, as is well known, every spin manifold gives rise to such data, 
          the above theorem demonstrates a one-to-one correspondence
 (up to unitary equivalence and spin structure preserving diffeomorphisms) 
between spin structures on spin manifolds 
          and real commutative geometries.
          \bigskip
\subsection{Proof of 1) and 2)}
           %Connected compact metrisable Hausdorff
Without loss of generality, we will make the simplifying assumption that 
$\pi$ is faithful on $\A$. This allows us to identify $\A$ with $\pi(\A)\subset\B(\HH)$, and 
we will simply write $\A$. As $\pi$ is a $^{*}$-homomorphism, the norm closure, 
$\overline{\A}$, is a 
          $C^{*}$-subalgebra of $\B(\HH)$.  
          Then the Gelfand-Naimark theorem tells us that $X=Spec(\overline{\A})$ is a compact, 
          Hausdorff space. Since $\A$ is dense in its norm closure, each state on $\A$ (defined with
respect to the $C^*$ norm of $\A$)
          extends to 
          a state on the closure, by continuity. Recall that we are assuming that $\A$ is
imbued with a norm such that the $C^*$-condition is satisfied for elements of $\A$; thus here
we mean continuity in the norm. Hence $Spec(\A)=Spec(\overline{\A})$. 
          The connectivity 
          of such a space is equivalent to the 
          non-existence of nontrivial projections in 
          $\overline{\A}\cong C(X)$. So let 
          $p\in\overline{\A}$ be such that $p^{2}=p$. Then 
          \be [\D,p]=[\D,p^{2}]=p[\D,p]+[\D,p]p=2p[\D,p].\ee
          So $(1-2p)[\D,p]=0$ implying that $[\D,p]=0$. By the irreducibility of $\pi$, 
          we must have $p=1$ or $p=0$. Hence $\A$ contains no non-trivial projections and $X$ 
          is connected. Note that the irreducibility also implies that 
          $[\D,a]\neq 0$ unless $a$ is a scalar. Also, as there are no 
          projections, any self-adjoint element of $\A$ has only continuous spectrum.
          \smallskip\newline
          The reader will easily show that equation (\ref{met}) does define a metric on $X$, \cite{MR}. The
topology defined by this metric is finer than the weak$^*$ topology, so functions continuous
for the weak$^*$ topology are automatically continuous for the metric. Furthermore, elements
of $\A$ are also Lipschitz, since for any $a\in\A$, $\n[\D,a]\n\leq 1$, we have 
$|a(x)-a(y)|\leq d(x,y)$. Thus for any $a\in\A$ we have  $|a(x)-a(y)|\leq\n[\D,a]\n d(x,y)$. Later we
will show that the metric and weak$^*$ topologies actually agree. This will follow from the fact that
$\A$, and so $\overline{\A}$, is finitely generated. This also  implies the separability of 
$C(X)\cong\overline{\A}$, which is equivalent to the metrizability of $X$. This will complete the
proof of $1)$ and $2)$, but it will
have to wait until we have learned some more about $\A$. The  last point of $2)$ is that the 
metric is invariant under unitary transformations. That is if 
         $U:\HH\rightarrow\HH$ is unitary
         \be [U\D U^{*},UaU^{*}]=U[\D,a]U^{*}\Rightarrow  \parallel[U\D U^{*},UaU^{*}]\parallel
=\parallel[\D,a]\parallel.\ee 
         So while $\D$ will be changed by a unitary change of representation 
\be \D:=\D_\pi\rightarrow U\D U^*:=\D_{U\pi U^*}=\D_\pi +U[\D_\pi,U^*],\ee
commutators with $\D$ change simply. For this reason, when we only need the unitary 
equivalence class 
of $\pi$, we drop the $\pi$, and write $\Omega^*_{\D}(\A)$ for $\pi(\Omega^*(\A))$, where 
the $\D$ 
is there to remind us that this is the representation of $\Omega^*(\A)$ induced by $\D$ and the 
first order condition. 

\subsection{Proof of 3) and remainder of 1) and 2)}
Before beginning the proof of $3)$, which will also complete the proof of $1)$ and $2)$, let us 
outline our approach, as this is the longest, and most important, portion of the proof.

\ref{a} {\bf Generalities}. This section deals with the various bundles involved, their Hermitian
structures and their relationships.  We also analyse the structure of
$\Omega^*_\D(\A)$ and $\Lambda^*_\D(\A)$, particularly in relation to Hochschild homology.

\ref{b} {\bf $X$ is a $p$-dimensional topological manifold}. We show that the 
elements of $\A$ involved in the Hochschild cycle
\be \pi(c)=\Gamma=\sum_ia^i_0[\D,a^i_1]\cdots[\D,a^i_p]\ee
 provided by the axioms generate $\A$. This is done in two steps. Results from \ref{a} show that
$\Gamma$ is antisymmetric in the $[\D,a^i_j]$, and this is used to show that
$\Omega^1_\D(\A)$ is finitely generated by the $[\D,a^i_j]$ appearing in $\Gamma$. The second
step then uses the long exact sequence in Hochschild homology to show that $\A$ is finitely
generated by the $a^i_j$ and $1\in\A$. In the process we will also show that $X$ is a topological
manifold.

\ref{c} {\bf $X$ is a smooth manifold}.
We show here that $\A$ is $C^\infty(X)$, and in particular that $X$ is a smooth
manifold. After proving that the weak$^*$ and metric topologies on $X$ agree, we show that 
$\A$ is closed under the holomorphic functional calculus, so that the
$K$-theories of $\A$ and $C(X)$ agree. At this point we will have completed the proof of $1)$ and
$2)$.

\ref{d} {\bf $X$ is a spin$^c$ manifold}. The form of the operators $\D$, $\dd$ and $\D^2$ 
is investigated. The main
result is that $\D^2$ is a generalised Laplacian while $\D$ is a generalised Dirac operator, in the
sense of \cite{BGV}. This allows us to show that $\Omega^*_\D(\A)$ is (at least
locally) the Clifford algebra of the complexified cotangent bundle. This is sufficient to 
show that the metric
given by equation (\ref{met}) agrees with the geodesic distance on $X$. As the representation of
$\Omega^*_\D(\A)$ is irreducible, we will have
completed the proof that $X$ is a spin$^c$ manifold.

\ref{e} {\bf $X$ is spin}. It is at this point that we utilise the real structure. Furthermore, we
reformulate Connes' result to allow a representation of the Clifford algebra of an indefinite metric.
This will necessarily involve a change in the underlying topology, which we do not investigate here.

\subsubsection{Generalities.}\label{a}
The axiom of finiteness and absolute continuity tells us that 
          $\HH_\infty=\cap_{m\geq 1}Dom\D^m$ is a finite projective $\A$ module. This tells us that 
          $\HH_\infty\cong e\A^N$, as an $\A$ module, for some $N$ and some $e=e^2\in M_N(\A)$.
Furthermore, from what we know about $\A$ and $Spec(\A)$,  $\HH_\infty$  is also 
isomorphic to a bundle 
of sections of a vector bundle over $X$, say $\HH_\infty\cong\Gamma(X,S)$. These sections 
will be of 
some degree of regularity which is at least continuous as $\A\subset C(X)$. This bundle is also 
imbued with 
an Hermitian structure $(\cdot,\cdot)_E:\HH_\infty\times\HH_\infty\rightarrow\A$ such that 
$(a\psi,b\eta)_S=a^*(\psi,\eta)_Sb$ etc, which provides us with an interpretation of the Hilbert space as 
$\HH=L^2(X,S,\intcross(\cdot,\cdot)|\D|^{-p})$. We will return to the important consequences of the 
Hermitian 
structure and the measure theoretic niceties of the above interpretation later.
\smallskip\newline
As mentioned earlier, $\A$ is a Frechet space for the locally convex topology coming from the 
family of seminorms
\be \n a\n,\ \ \n\delta(a)\n,\ \ \n\delta^2(a)\n,...\ \ \ \forall a\in\A,\ \ \delta(a)=[\dd,a].\ee
Note that the first semi-norm in this family is the $C^*$-norm of $\A$, and that $\delta^n(a)$ makes
sense for all $a\in\A$ by hypothesis. As the first semi-norm is in fact a norm, this topology is 
Hausdorff. In fact our hypotheses allow us to extend these seminorms
to all of $\Omega^*_\D(\A)$, and it too will be complete for this topology.
\smallskip\newline
Now let us turn to the differential structure. The first things to note are that 
$\D\HH_\infty\subset\HH_\infty$, $\A\HH_\infty\subset\HH_\infty$ and $\n[\D,a]\n<\infty$ 
$\forall a\in\A$. The associative algebra $\Omega^*_{\D}(\A)$ is generated by $\A$ and 
$[\D,\A]$, so $\Omega^*_{\D}(\A)\HH_\infty\subset\HH_\infty$. In other words
\be \Omega^*_{\D}(\A)\subset End(\Gamma(X,S))\cong End(e\A^N)\cong
\{B\in M_N(\A): Be=eB\}.\ee
The most important conclusion of these observations is that $\Omega^*_{\D}(\A)$ and so  
$\Omega^1_{\D}(\A)$ are finite projective over $\A$, and so are both (sections of) vector bundles 
over $X$, the 
former being (the sections of) a bundle of algebras as well. Moreover, the 
irreducibility hypothesis tells us that 
$\Gamma(X,S)$ is an irreducible module (of sections) for the algebra (of sections) 
$\Omega^*_{\D}(\A)$.
\smallskip\newline
The finite generation of $\Omega^*_\D(\A)$ can be used to show that the algebra $\A$ is finitely
generated. We will then be dealing with a finitely
generated algebra and this can be used to show that the weak$^*$ and metric topologies agree. When we
have completed the proofs of these matters, 1) will have been proved completely. 
\smallskip\newline
So what is $\Omega^*_{\D}(\A)$? The central idea for studying this algebra is the 
first 
order condition. When we construct this representation of $\Omega^*(\A)$ 
from $\pi$ and $\A$ using $\D$, the first order condition 
forces us to identify the left and right actions of $\A$ on $\Omega^*_{\D}(\A)$, at least in the 
commutative case. Assuming as we 
are that the representation is faithful on $\A$, we see that the ideal $\ker\pi$ is generated by the 
first order condition,
\be \ker\pi = \langle \omega a-a\omega\rangle_{a\in\A,\omega\in\Omega^*(\A)}=
\langle \mbox{first order condition}\rangle.\label{hochs}\ee
So for $\omega=\delta f$ of degree $1$ and $a\in\A$, $a\delta f-\delta f a\in\ker\pi$ and
\be  (\delta f)(\delta a)+(\delta a)(\delta f)\in\delta\ker\pi.\label{precliff}\ee
Equation (\ref{hochs}) ensures that $\pi\circ b=0$, as $\mbox{Image}(b)=\ker\pi$, so that we have a well-defined
faithful representation of Hochschild homology
\be \pi:HH_*(\A)\to\Omega^*_{\D}(\A).\ee 
If we write $d=[\D,\cdot]$, 
we see that the existence of junk is due to the fact that 
$\pi\circ\delta\neq d\circ\pi$, and that this may be traced directly to the first order condition. 
Let us continue to 
write $\Omega^*_{\D}(\A):=\pi(\Omega^*(\A))$ and also write 
$\Lambda^*_{\D}(\A):=\pi(\Omega^*(\A))/\pi(\delta\ker\pi)$, and note that the second algebra is 
skew-commutative, and a graded differential algebra for the differential $d=[\D,\cdot]$. Note that this 
notation differs somewhat from the usual, \cite{C}. Note that $\Omega^1_\D(\A)$ and
$\Lambda^1_\D(\A)$ are the same finite projective $\A$ module, and we denote them both by
$\Gamma(X,E)$ for some bundle $E\to X$, where as before we do not specify the regularity of the sections, only that they are at
least continuous for the weak$^*$ topology and Lipschitz for the metric topology.
\smallskip\newline
The next point to examine is $\delta\ker\pi=\mbox{Image}(\delta\circ b)$. This is easily seen to be generated
by $\ker\pi=\mbox{Image}(b)$ and graded commutators $(\delta
a)\omega-(-1)^{|\omega|}\omega\delta a$,
for $\omega\in\Omega^*(\A)$ and $a\in\A$. Thus the image of $\pi\circ\delta\circ b$ in
$\Omega^*_\D(\A)$ is junk, and this is graded commutators. As elements of the form appearing in
equation (\ref{precliff}) generate $\pi(\delta\ker\pi)$, it is useful to think of 
equation (\ref{precliff}) as a kind of `pre-Clifford' relation. In particular, controlling the
representation of elements of $\delta\ker\pi$ will give rise to a representation of the Clifford
algebra as well as the components of the metric tensor. More on that later.
\smallskip\newline
To help our analysis, define $\s:\Omega^*(\A)\to\Omega^*(\A)$ by $\s(a)=a$ for
all $a\in\A$ and $\s(\omega\delta a)=(-1)^{|\omega|}(\delta a)\omega$, for $|\omega|\geq 0$.
Then, \cite{L}, we have $b\circ\delta+\delta\circ b=1-\s$, on $\Omega^*(\A)$. As $\pi\circ b=0$, and 
$\mbox{Image}(\pi\circ\delta\circ
b)=$Junk, we have $\mbox{Image}(\pi\circ (1-\s))=$Junk. 
\smallskip\newline
Thus while
\be\pi(\Omega^*(\A))=\Omega^*_\D(\A)\cong\Omega^*(\A)/\langle\mbox{Image}(b)\rangle,\ee
passing to the junk-free situation gives
\be \Omega^*(\A)/\langle\mbox{Image}(b),\mbox{Image}(1-\s)\rangle\cong\Lambda^*_\D(\A)
\cong\widehat{\Omega}^*(\A).\ee
It is easy to see that $b(1-\s)=(1-\s)b$, so that $\ker b$ is preserved by $1-\s$. In fact, $1-\s$ sends
Hochschild cycles to Hochschild boundaries. For if $bc=0$ for some element $c\in C_n(\A)$, then 
\be (1-\s)c=(b\delta+\delta b)c=b\delta c\ee
which is a boundary. So $\ker b$ is mapped into Image$(b)$ under $1-\s$ and so when we quotient by
Image$(1-\s)$ we do not lose any Hochschild cycles. 
\smallskip\newline
So, $\pi$ descends to a
faithful representation of Hochschild homology with values in $\Lambda^*_\D(\A)$. In general, the 
Hochschild homology groups of a commutative and unital algebra contain 
$\widehat{\Omega}^*(\A)$ as a direct summand, \cite{L}, but we have shown that in fact 
\be HH_*(\A)\cong\Lambda^*_\D(\A)\cong\widehat{\Omega^*}(\A).\ee
This is certainly a necessary condition for the algebra $\A$ to be smooth, but more important for us at
this point is that all Hochschild cycles are antisymmetric in elements of $\Omega^1_\D(\A)$. In
particular, $\pi(c)=\Gamma\neq 0$ in $\Lambda^p_\D(\A)$ and is totally antisymmetric. 
\smallskip\newline
There is another consequence of this result. 
If we define an Hermitian form on $\Omega^1_\D(\A)$ by
$(da,db)_{\Omega^1}=\frac{1}{M}Trace((da)^*db)$, then 
\be (da\xi,db\eta)_S=(da,db)_{\Omega^1}(\xi,\eta)_S,\ee
where $M$ is the fibre dimension of the bundle underlying $\Omega^1_\D(\A)$, 
and this uses the antisymmetry in an essential way.  It also shows that the trace is a nondegenerate
Hermitian form on $\Omega^1_\D(\A)$.
\subsubsection{$X$ is a $p$-dimensional topological manifold.}\label{b} 
We claim that the elements $a^i_j$, $i=1,...,n$ $j=1,...,p$ appearing in the Hochschild cycle
$\Gamma$, along with $1\in \A$, generate $\A$ as an
algebra over ${\C}$. Without loss of generality we take $a^i_j$ to be self-adjoint for $i,j\geq 1$.
Furthermore, we may also assume that $\n[\D,a^i_j]\n=1$.
To show that the $a^i_j$ generate, we first show that the $[\D,a^i_j]$ generate $\Omega^1_\D(\A)$. Let $\Gamma$ be
the (totally antisymmetric) representative of the Hochschild $p$-cycle provided by the axioms. We
write $da:=[\D,a]$ for brevity, and similarly we write $d$ for the action of $[\D,\cdot]$ on forms.
\smallskip\newline
Recall that $\Gamma=\pi(c)$, and note that for any $a\in\A$
\bea \pi(1-\s)(c\delta a)=\Gamma da-(-1)^pda\Gamma\qquad\quad\ \nno
=(1-(-1)^p(-1)^{p-1})\Gamma da\nno
=2\Gamma da.\qquad\qquad\qquad\quad\ \ \eea
Thus $\Gamma da$ is junk and so contains a symmetric factor, and $da\wedge\Gamma=0$ for all $a\in\A$. 
In order to show that the $da^i_j$ generate
$\Omega^1_\D(\A)$ as an $\A$ bimodule, we need to show that $da\wedge\Gamma=0$ implies that
$da$ is a linear combination of the $da^i_j$. To do this, first write 
\be \Gamma=\sum_{i=1}^na^i_0da^i_1\cdots da^i_p=\sum_{i=1}^n\Gamma^i.\ee
Now suppose that $n=1$, so that $\Gamma=a_0da_1\cdots da_p$. Then if 
\be (da\wedge\Gamma)(x)=(a_0da\wedge da_1\cdots da_p)(x)=0\ee
for all $x\in X$, elementary exterior algebra tells us that $da(x)$ is a linear combination of
$da_1(x),..., da_p(x)$ in each fibre. 
\smallskip\newline
So if $n>1$, the only thing we need to worry about is cancellation in the sum
\be \sum da\wedge\Gamma^i.\ee
Without loss of generality, we can assume that at each $x\in X$ there is no cancellation in the sum 
\be \sum\Gamma^i.\ee
So for all $I,J\subset\{1,...,n\}$ with $I\cap J=\emptyset$, 
\be \sum_{i\in I}\Gamma^i(x)\neq -\sum_{j\in J}\Gamma^j(x).\ee
If there were such terms we could simply remove them anyway, and we know in doing so we do not
remove all the terms $\Gamma^i$ as $\Gamma(x)\neq 0$ for all $x\in X$.
\smallskip\newline
Now suppose that for some $x\in X$ and some $I,J\subset\{1,...,n\}$ with $I\cap J=\emptyset$ we
have 
\be (\sum_{i\in I}da\wedge\Gamma^i)(x)=-(\sum_{j\in J}da\wedge\Gamma^j)(x).\ee
If $da(x)$ is a linear combination of any of the terms appearing in these $\Gamma^i$'s, we are done.
So supposing that $da$ is linearly independent of the terms appearing in 
$\sum_{I\cup J}\Gamma^i$, we have
\be \sum_{i\in I}\Gamma^i(x)=-\sum_{j\in J}\Gamma^j(x)\ee
contradicting our assumption on $\Gamma$. Thus we may assume that no terms cancel, which
shows that
\be (da\wedge\Gamma^i)(x)=0\ee
for each $i=1,...,n$. Hence if $\Gamma^i(x)\neq 0$, $da(x)$ is linearly dependent on $da^i_1(x),...,da^i_p(x)$. This also shows that if $\Gamma^i,\Gamma^j$ are both nonzero at
$x\in X$, then they are linearly dependent at $x$.
These considerations show that the $da^i_j$ generate $\Omega^1_\D(\A)$ as an $\A$ bimodule.
\smallskip\newline
As a consequence, the $da^i_j$ also generate $\Lambda^*_\D(\A)$ as a graded differential algebra.
From what we have already shown, this algebra is precisely 
\be \Lambda^*_\A\Omega^1_\D(\A)=\Lambda^*_\A\Gamma(E)=\Gamma(\Lambda^*E).\ee
Now the $da^i_j$ generate and any $p+1$ form in them is zero from the above argument, while we
know that $\Lambda^p_\D(\A)\neq\{0\}$ because $\Gamma\in\Lambda^p_\D(\A)$. Also, for all $x\in
X$, we know that $\Gamma(x)\neq 0$, so each fibre $\Lambda^pE_x$ is nontrivial. Lastly, using
the antisymmetry and non-vanishing of $\Gamma$, it is easy to see that for all $x\in X$ there is an $i$ such that the
$da^i_j(x)$, $j=1,...,p$, are linearly independent in $E_x$. For if , say, 
\be da^i_1(x)=\sum_{j=2}^pc_jda^i_j(x)\ee
then inserting this expression into the formula for $\Gamma$ and using the antisymmetry shows that
\be da^i_1da^i_2\cdots da^i_p(x)=0.\ee
If this happenned for all $i$ at some $x\in X$ we would have a contradiction of the non-vanishing of
$\Gamma(x)$. Hence we can always find such an $i$. 
\smallskip\newline
Putting all these facts together, and
recalling that $X$ is connected, we see that $E$ has rank $p$ as a vector bundle, and moreover,
for all $x\in X$ there is an index $i$ such that the $da^i_j(x)$ form a basis of $E_x$. Later we will see
that $E$ is essentially the (complexified) cotangent bundle. 
\smallskip\newline
We now have the pieces necessary to show that $\A$ is in fact finitely 
generated by the $a^i_j$. 
Suppose that the functions $a^i_j$ do not separate the points of $X$. Define an equivalence relation
on $X$ by
\be x\sim y\Leftrightarrow a^i_j(x)=a^i_j(y)\ \ \forall i,j.\ee
Then by adding constants to the $a^i_j$ if necessary, there is an equivalence class $B$ such that
\be a^i_j(B)=\{0\}\ \ \forall i,j.\ee
So the $a^i_j$ generate an ideal $\langle a^i_j\rangle$ whose norm closure is 
$C_0(X\setminus B)$. The fact that $\Lambda^*_\D(\A)$ is complete in the topology determined by the family of
seminorms provided by $\delta$, and is a locally convex
Hausdorff space for this topology, shows that the topological Hochschild homology is Hausdorff, and
allows us to use the long exact sequence in topological Hochschild homology.  
We have the exact sequence 
\be 0\to\langle a^i_j\rangle\stackrel{I}{\to}\A\stackrel{P}{\to}\A/\langle a^i_j\rangle\to 0\ee
as well as a norm closed version
\be 0\to C_0(X\setminus B)\to C(X)\to C(B)\to 0\ee
The former sequence, being a sequence of locally convex algebras, induces a long exact sequence in
topological Hochschild homology. The bottom end of this looks like
\be\cdots\to \Lambda^2(\A/\langle a^i_j\rangle)\to\Lambda^1_\D(\langle
a^i_j\rangle)\to\Lambda^1_\D(\A)\to\Lambda^1_\D(\A/\langle a^i_j\rangle)\to\langle a^i_j\rangle
\stackrel{I}{\to}\A\stackrel{P}{\to}\A/\langle a^i_j\rangle\to 0.\ee
From what we have shown, every element of $\Lambda^n_\D(\A)$ is of the form
\be \omega=\sum(a\hat{\otimes}a_1\hat{\otimes}a_2\hat{\otimes}\cdots\hat{\otimes}a_n)\ee
where $a\in\A$ and $a_k\in\langle a^i_j\rangle$ for each $1\leq k\leq n$. The map induced on homology
by $P:\A\to\A/\langle a^i_j\rangle$ is easy to compute:
\be P_*\sum(a\hat{\otimes}a_1\hat{\otimes}\cdots\hat{\otimes}a_n)=
\sum(P(a)\hat{\otimes}P(a_1)\hat{\otimes}\cdots\hat{\otimes}P(a_n))=
\sum(P(a)\hat{\otimes}0\cdots\hat{\otimes}0)=0.\ee
\smallskip\newline
So $\Lambda^n_\D(\A/\langle a^i_j\rangle)=0$ for all $n\geq 1$. The case $n=1$ says that 
\be \delta P(a)=1\otimes P(a)-P(a)\otimes 1=0\Rightarrow P(a)\in{\C}\cdot 1.\ee
Hence $C(B)={\C}$ and $B=\{pt\}$.
As an immediate corollary we see that all the equivalence classes of $\sim$ are singletons, so $\A$ is
generated in its Frechet topology by the elements $a^i_j$.
\smallskip\newline
Now take the natural open cover of $X$ given by the open sets
\be U^i=\{x\in X: [\D,a^i_1],...,[\D,a^i_p]\neq 0\}.\ee
From what we have already shown, over this open set we obtain a local trivialisation
\be E|_{U^i}\cong U^i\times {\C}^p.\ee
As 
\be |a^i_j(x)-a^i_j(y)|\leq\n[D,a^i_j]|_F\n d(x,y)\ee
where $F$ is any closed set containing $x$ and $y$, we see that the $a^i_j$ are constant off $U^i$. By
altering these functions by adding scalars, we see that we can take their value off $U^i$ to be zero.
Thus $\langle a^i_j\rangle_j\subseteq C_0(U^i)$. Noting that the $da^i_j$ provide a generating set
for $\Omega^1_\D(\A_{U^i})$ over $\A_{U^i}$ (the closure of the functions in
$\A$ vanishing off $U^i$ for the Frechet topology), the previous argument shows that the 
$a^i_j$ generate $\A_{U^i}$ in the Frechet topology and $C_0(U^i)$ in norm. The inessential 
detail that $\A_{U^i}$ is not
unital may be repaired by taking the one point compactification of $U^i$ or simply noting that the
above argument runs as before, but now the only scalars are zero, whence the equivalence class $B$
is empty.
\smallskip\newline
We are now free to take as coordinate charts $(U^i,a^i)$ where $a^i=(a^i_1,...,a^i_p):U^i\to{\R}^p$.
As both the $a^i_j$ and the $a^k_j$ generate the functions on $U^i\cap U^k$, we may deduce the 
existence of continuous transition functions $f^i_{jk}:{\R}^p\to{\R}^p$ with compact support such that
\be a^i_j=f^i_{jk}(a^k_1,...,a^k_p)\mbox{ on the set }U^i\cap U^k.\ee
As these functions are necessarily continuous, we have shown that $X$ is a topological manifold, and
moreover the map $a=(a^1,...,a^n):X\to{\R}^{np}$ is a continuous embedding.

\subsubsection{$X$ is a smooth manifold}\label{c}
We can now show that $X$ is a smooth manifold. On the intersection $U^i\cap U^k$, the
functions can be taken to be generated by either $a^i_1,...,a^i_p$ or $a^k_1,...,a^k_p$. Thus we may
write the transition functions as
\be a^i_j=f^i_{jk}=\sum^\infty_{N=0}p_N(a^k_j)\ee
where the $p_N$ are homogenous polynomials of total degree $N$ in the $a^k_j$. As the $a^i_j$
generate $\A$ in its Frechet topology, we may assume that this sum is
convergent for all the seminorms $\n\delta^n(\cdot)\n$.
Also, $\Omega^*_\D(\A)\subset\cap_{n\geq 1}Dom\delta^n$ and
$[\D,\cdot]:\Omega^*_\D(\A)\to\Omega^*_\D(\A)$, showing that the sequence
\be \sum_{N=0}^\infty[\D,p_N]\ee
converges. Since $\D$ is a closed operator, the derivation $[\D,\cdot]$ can be seen to be closed as
well. Thus, over the open set $U^i\cap U^k$, we see that the above sequence converges to $[\D,a^i_j]$, so
\be [\D,a^i_j]=\sum_{l=1}^p\sum_{N=0}^\infty\frac{\p p_N}{\p a^k_l}[\D,a^k_l],\ee
where we have also used the first order condition. Consequently, the functions
\be \sum_{N=0}^\infty\frac{\p p_N}{\p a^k_l}\in\A\subset C(X)\ee
are necessarily continuous. This allows us to identify
\be \frac{\p f^i_{jk}}{\p a^k_l}=\sum_{N=0}^\infty\frac{\p p_N}{\p a^k_l}.\ee
Applying the above argument repeatedly to the functions 
$\frac{\p f^i_{jk}}{\p a^k_l}$, shows that $f^i_{jk}$ is a $C^\infty$ function. Hence $X$ is a smooth
manifold for the metric topology and $\A\subseteq C^\infty(X)$. In particular, the functions $a^i_j$
are smooth. 
\smallskip\newline
Conversely, let $f\in C^\infty(X)$. Over any open set $V\subset U^i$ we may write 
\be f=\sum_{n=0}^\infty p_N(a_1,..,a_p)\ee
where we have temporarily written  $a_j:=a^i_j$. As $f$ is smooth, all the sequences
\be \sum_{|\alpha|=n}\frac{\p^{|\alpha|}f}{\p^{\alpha_1}a_1\cdots\p^{\alpha_p}a_p}=
\sum_{|\alpha|=n}\sum_{N=0}^\infty\frac{\p^{|\alpha|}p_N}{\p^{\alpha_1}a_1\cdots
\p^{\alpha_p}a_p}\ee
converge, where $\alpha\in{\N}^n$ is a multi-index. Let $p_N=\sum_{|\alpha|=N}C_\alpha
a_1^{\alpha_1}\cdots a_p^{\alpha_p}$ and let $s_M=\sum_{N=0}^M p_N$ be the partial sum. Then
\bea
[\dd,s_M]=\sum_{N=0}^M\sum_{|\alpha|=N}\sum_{j=1}^p\sum_{k=1}^{n_j}C_Na_1^{n_1}\cdots
a_j^{n_j-k}[\dd,a_j]a_j^{k-1}\cdots a_p^{n_p}\quad\nno
=\sum_{n=0}^M\sum_{|\alpha|=N}\sum_{j=1}^p\sum_{k=1}^{n_j}C_\alpha a_1^{\alpha_1}\cdots 
a_j^{\alpha_j-1}\cdots
a_p^{\alpha_p}[\dd,a_j]\qquad\quad\ \nno
+\sum_{n=0}^M\sum_{|\alpha|=N}\sum_{j=1}^p\sum_{k=1}^{n_j}C_\alpha a_1^{\alpha_1}\cdots
a_j^{\alpha_j-k}[[\dd,a_j],a_j^{k-1}\cdots a_p^{\alpha_p}]\ \nno
=G^1_M+\sum_{j=1}^p\frac{\p s_M}{\p
a_j}[\dd,a_j].\qquad\qquad\qquad\qquad\qquad\quad\qquad\eea
To show that $f\in Dom\delta$, we must show that $G^1_M$ can be bounded independent of $M$, the
other term being convergent by the smoothness of $f$ and the boundedness of $[\dd,a_j]$ for each
$j$. We have the following bound
\bea \n G^1_M\n\leq \sum_{N=0}^M\sum_{|\alpha|=N}\sum_{j=1}^p\sum_{k=1}^{n_j}\n
C_\alpha a_1^{\alpha_1}\cdots
a_j^{\alpha_j-k}[[\dd,a_j],a_j^{k-1}\cdots a_p^{\alpha_p}]\n\qquad\qquad\ \ \nno
\leq \sum_{N=0}^M\sum_{|\alpha|=N}\sum_{j=1}^p\sum_{k=1}^{n_j} 2|C_N|\n a_1\n^{n_1}\cdots\n a_j\n^{n_j-1}\cdots\n
a_p\n^{n_p}\n[\dd,a_j]\n\nno
=2\sum_{N=0}^M\sum_{j=1}^p\frac{\p \tilde{p}_N}{\p a_j}(\n a_1\n,...,\n
a_p\n)\n[\dd,a_j]\n,\qquad\qquad\qquad\qquad\qquad\quad\eea
where $\tilde{p}_N=\sum_{|\alpha|=N}|C_\alpha|a_1^{\alpha_1}\cdots a_p^{\alpha_p}$. The
absolute convergence of the sequence of real numbers
\be \sum_{N=0}^\infty\frac{\p p_N}{\p a_j}(\n a_1\n,...,\n a_p\n)\ee
now shows that $\n G^1_M\n$ can be bounded independtly of $M$. Thus the sequence $[\dd,s_M]$
converges, and as $[\dd,\cdot]$ is a closed derivation, it converges to $[\dd,f]$. Hence
$\n[\dd,f]\n<\infty$, and $f\in Dom\delta$. 
\smallskip\newline
Applying $\delta$ twice gives
\be
[\dd,[\dd,s_M]]=G^2_M+\sum_{|\alpha|=2}
\frac{\p^2 s_M}{\p^{\alpha_j}a_j\p^{\alpha_k}a_k}[\dd,a_j]^{\alpha_j}[\dd,a_k]^{\alpha_k}+\sum_{|\alpha|=1}\frac{\p s_M}{\p
a_j}\delta^2(a_j).\ee
The second two terms can be bounded independently of $M$ by the smoothness of $f$. The term
$G^2_M$ is a sum of commutators and double commutators which can be bounded independently of
$M$ in exactly the same manner as $G^1_M$. This shows that
\be \n[\dd,[\dd,f]]\n<\infty\ee
and $f\in Dom\delta^2$. Continuing this line of argument shows that $f\in Dom\delta^n$ for all $n$,
and so $f\in\A$. Consequently, $\A=C^\infty(X)$, and the seminorms $\n\delta^n(\cdot)\n$
determine the $C^\infty$ topology on $\A$.
\bigskip\newline
To show that the weak$^*$ and metric topologies agree, it is sufficient to show that convergence in the
weak$^*$ topology implies convergence in the metric topology, as the metric topology is
automatically finer. 
\smallskip\newline
So let $\{\phi_k\}_{k=1}^\infty$ be a weak$^*$ convergent sequence of pure states of $\A$ (or
$\overline{\A}$). Thus there is a pure state $\phi$ such that for all $f\in\A$,
\be |\phi_k(f)-\phi(f)|\to 0.\ee
As $\A$ is commutative, we know that every pure state is a $*$-homomorphism, and writing the
generating set of $\A$ as $a_1,...,a_{np}$ we have for $f=\sum p_N$, 
\be \phi_k(f)=\sum^\infty_{N=0}p_N(\phi_k(a_i))\ee
and this makes sense since the sum is convergent in norm.
\smallskip\newline
The next aspect to address is the norm of $[\D,f]$. Recalling that $\n[\D,a_i]\n=1$, we have 
\bea \n[\D,f]\n^2=\n\sum_{i,j=1}^{np}\frac{\p f}{\p a_i}[\D,a_i](\frac{\p f}{\p
a_j})^*[\D,a_j]^*\n\qquad\ \ \nno
\leq\sup_{x\in X}|\sum_{i,j=1}^{np}\frac{\p f}{\p a_i}(x)(\frac{\p f}{\p a_j})^*(x)|\qquad\qquad\ \ \nno
\qquad\qquad\qquad\qquad=\sup_{a(x)\in a(X)}|\sum_{i,j=1}^{np}\frac{\p f}{\p x_i}(a(x))(\frac{\p f}{\p x_j})^*(a(x))|\nno
=\n df\n^2,\ \mbox{regarding }f:{\R}^{np}\to{\C},\qquad\ \ \eea
where $a:X\to{\R}^{np}$ is our (smooth) embedding and $x_i$ are coordinates on ${\R}^{np}$. 
Thus $\n df\n\leq 1\Rightarrow\n[\D,f]\n\leq 1$. Any function $f:{\R}^{np}\to{\C}$ satisfying $\n
df\n\leq 1$ is automatically Lipschitz (as a function on ${\R}^{np}$). So
\bea |\phi_k(f)-\phi(f)|=|f(\phi_k(a_i))-f(\phi(a_i))|\qquad\qquad\ \nno
\qquad\qquad\leq |\phi_k(a_i)-\phi(a_i)|\to 0\mbox{ as }k\to\infty.\eea
Hence 
\bea\sup\{|\phi_k(f)-\phi(f)|:\n[\D,f]\n\leq 1\}=\sup\{|f(\phi_k(a_i))-f(\phi(a_i))|:\n df\n\leq
1\}\ \nno
\leq \{|\phi_k(a_i)-\phi(a_i)|\}\to 0\qquad\qquad\qquad\quad\eea
so $\phi_k\to\phi$ in the metric. So the two topologies agree.
\bigskip\newline
As a last note on these issues, it is important to point out that $\A$ is stable under the holomorphic
functional calculus. If $f:X\to{\C}$ is in $\A$, then we may (locally) regard it as a smooth function
$f:{\R}^p\to{\C}$ of $a^i_1,...,a^i_p$ for some $i$. So let $g:{\C}\to{\C}$ be holomorphic. Then 
\be g\circ f\circ a^i\ee
is patently a smooth function on $X$. Thus the $K$-theory and $K$-homology of $\A$ and $\overline{\A}$
coincide, \cite{C}.

\subsubsection{$X$ is a spin$^c$ manifold}\label{d}
We have been given an Hermitian structure on $\HH_\infty$, $(\cdot,\cdot)_S$, and as
$\Omega^1_\D(\A)$ is finite projective, we are free to choose one for it also. Regarding
$\Omega^*_\D(\A)$ as a subalgebra of $End(\HH_\infty)$, any non-degenerate Hermitian form we choose is unitarily
equivalent to $([\D,a],[\D,b])_{\Omega^1}:=\frac{1}{p}Tr([\D,a]^*[\D,b])$, where $p$ is the fibre
dimension of $\Omega^1_\D(\A)$. We have shown this is a non-degenerate positive definite quadratic 
form.  Over each $U^i$, we have a local trivialisation  (recalling that we have
set $\Omega^1_\D(\A)=\Gamma(X,E)$)
\be E|_{U^i}\cong U^i\times{\C}^p.\ee
As $X$ is a smooth manifold, we can also define the cotangent bundle, and as the $a^i$ are local
coordinates on each $U^i$, we have
\be T^*_{\C}X|_{U^i}\cong U^i\times{\C}^p.\ee
It is now easy to see that these bundles are locally isomorphic. Globally they may not be
isomorphic, though. The reason is that while we may choose $T^*_{\C}X$ to be $T^*X\otimes{\C}$
globally, we do not know that this is true for $\Omega^1_\D(\A)$. Nonetheless, up to a possible
$U(1)$ twisting, they are globally isomorphic. It is easy to show using our change of coordinate
functions $f^i_{jk}$ that up to this possible phase factor the two bundles have the same transition
functions. For the next step of the proof we require only local
information, so this will not affect us. Later we will use the real structure to show that
$\Omega^*_\D(\A)$ is actually untwisted. 
\smallskip\newline
From the above comments, we may easily deduce that
\be \Lambda^*_\D(\A)|_{U^i}\cong \Gamma(\Lambda^*_{\C}(T^*X))|_{U^i}.\ee
The action of $d=[\D,\cdot]$ on this bundle may be locally determined, since we know that
$\Lambda^*_\D(\A)$ is a skew-symmetric graded differential algebra for $d$. First $d^2=0$, and
$d$ satisfies a graded Liebnitz rule on $\Lambda^*_\D(\A)$. Furthermore, from the above local
isomorphisms, given $f\in\A$, 
\be df|_{U^i}=\sum_{j=1}^p\frac{\p f}{\p a^i_j}[\D,a^i_j]=\sum_{j=1}^p\frac{\p f}{\p
a^i_j}da^i_j.\ee
By the uniqueness of the exterior derivative, characterised by these three properties, $[\D,\cdot]$ is
the exterior derivative on forms. We shall continue to write $d$ or $[\D,\cdot]$ as convenient. 
\smallskip\newline
Let us choose a connection compatible with the form $(\cdot,\cdot)_S$ 
\be \nabla:\HH_\infty\to\Lambda^1_\D(\A)\otimes\HH_\infty\ee 
\be \nabla(a\xi)=[\D,a]\otimes\xi+a\nabla\xi.\ee
Note that from the above discussion, this notion of connection agrees with our usual idea of covariant
derivative. Denote by $c$ the obvious map
\be c:End(\HH_\infty)\otimes\HH_\infty\to \HH_\infty\ee
and consider the composite map $c\circ\nabla:\HH_\infty\to\HH_\infty.$ We have 
\bea (c\circ\nabla)(a\xi)=[\D,a]\xi+c(a\nabla\xi),\ \ \forall a\in\A,\ \xi\in\HH_\infty\nno
=[\D,a]\xi+ac(\nabla\xi)\qquad\qquad\qquad\qquad\ \eea
whereas
\be \D(a\xi)=[\D,a]\xi+a\D\xi\ \ \ \forall a\in\A,\ \xi\in\HH_\infty.\ee
Hence, on $\HH_\infty$, 
\be (c\circ\nabla-\D)(a\xi)=a(c\circ\nabla-\D)\xi\ee
so that $c\circ\nabla-\D$ is $\A$-linear, or better, in the commutatant of $\A$. Thus if $c\circ\nabla-\D$ is bounded,
it is in the weak
closure of $\Omega^*_\D(\A)$. However, as $(c\circ\nabla-\D)\HH_\infty\subseteq\HH_\infty$, it must in fact be in
$\Omega^*_\D(\A)$. The point of these observations is that if $c\circ\nabla-\D$ is bounded, then as $\nabla$ is a first
order differential operator (in particular having terms of integral order only) so is $\D$ (as elements
of $\Omega^*_\D(\A)$ act as endomorphisms of $\HH_\infty$, and so are order zero operators.) 
So let us show that
$c\circ\nabla-\D$ is bounded. We know $\HH_\infty\cong e\A^N$ for some $N$ and $e\in M_N(\A)$. As both $\D$
and $\nabla$ have commutators with $e$ in $\Omega^*_\D(\A)$ 
(because $\D$, $c\circ\nabla:\HH_\infty\to\HH_\infty$) there is no loss of generality in setting $e$ to $1$ for our immediate purposes. So, simply
consider the canonical generating set of $\HH_\infty$ over $\A$ given by $\xi_j=(0,...,1,...,0)$, $j=1,...,N$. Then, there are
$b_i^j,c_i^j\in\A$ such that 
\be c\circ\nabla\xi_i=\sum_jb_i^j\xi_j,\quad \D\xi_i=\sum_jc_i^j\xi_j.\ee
As $c\circ\nabla-\D$ is $\A$-linear, this shows that $c\circ\nabla-\D$ is bounded. Hence $\D$ is a 
first order differential operator. As the difference $c\circ\nabla-\D$ is in $\Omega^*_\D(\A)$,
$c\circ\nabla-\D=A$, for some element of $\Omega^*_\D(\A)$. However, as 
$c\circ\nabla=\D+A$ is a connection (ignoring $c$), $A\in\Omega^1_\D(\A)$.
\smallskip\newline
Thus over $U^i$, we may write the matrix form of $\D$ as 
\be \D^k_{\ m}=\sum_{j=1}^p\alpha^{\ k}_{j\ m}\frac{\partial}{\partial a_j}+\beta^k_{\ m}\ee
where $\beta^k_{\ m}$, $\alpha^{\ k}_{j\ m}$ are bounded for each $k,m$. Similarly we write the square of $\D$ as
\be (\D^2)^{n}_{\ m}=\sum_{j,k}A_{jk\ m}^{\ \ n}\frac{\partial^2}{\partial a_j\partial a_k}+
\sum_kB^{\ n}_{k\ m}\frac{\partial}{\partial a_k}+C^n_{\ m}\ee
with all the terms $A,B,C$ bounded, so that (as a pseudodifferential operator)
\be \dd^n_{\ m}=\sum_kE_{k\ m}^{\ n}\frac{\partial}{\partial a_k}+F^n_{\ m}\ee
where $E,F$ are bounded and 
\be \sum_mE^{\ n}_{k\ m}E^{\ m}_{j\ p}=A_{kj\ p}^{\ \ n}\ee
et cetera. We will now show that the boundedness of $[\dd,[\D,a]]$, required by the axioms, tells us that
the first order part of $\dd$ has a coefficient of the form $fId_N$, for some $f\in\A$. 
With the above notation,
\be [\dd,[\D,a]]^n_{\ p}=\sum_{k,m}E^{\ n}_{k\ m}(\frac{\partial[\D,a]^m_{\ p}}{\partial a_k})+
\sum_{k,m}(E^{\ n}_{k\ m}[\D,a]^m_{\ p}-[\D,a]^n_{\ m}E^{\ m}_{k\ p})\frac{\partial}{\partial a_k}+
[F,[\D,a]]^n_{\ p}.\ee
For this to be bounded, it is necessary and sufficient that $[E_k,[\D,a_j]]=0$, for all $j,k=1,...,p$. As
\be [\dd,[\D,a_j][\D,a_k]]=[\D,a_j][\dd,[\D,a_k]]+[\dd,[\D,a_j]][\D,a_k],\ee
and the commutant of $\Omega^*_\D(\A)$ restricted to $U^i$ is the weak closure of $\A$ restricted to $U^i$, 
the matrix $E_k$ must be scalar over $\A$ for each $k$ (not $\A^{\prime\prime}$ since
$\dd\HH_\infty\subseteq\HH_\infty$). Thus $E^{\ n}_{k\ m}=f_k\delta^n_{\ m}$, for
some $f_k\in\A$. Since
\be A^{\ \ n}_{kj\ p}=f_kf_j\delta^n_{\ p}\ee
the leading order terms of $\D^2$ also have scalar coefficients.
\smallskip\newline
Using the first order condition we see that
\be [\D,a_j][\D,a_k]+[\D,a_k][\D,a_j]=[[\D^2,a_j],a_k]=[[\D^2,a_k],a_j],\label{clap}\ee
and denoting by
$g^i_{jk}:=([\D,a^i_j],[\D,a^i_k])_{\Omega^1}$, we have
\be \frac{1}{p}Tr([\D,a_j][\D,a_k]+[\D,a_k][\D,a_j])=-2Re(g^i_{jk}),\ee
since $[\D,a_j]^*=-[\D,a_j]$. Now (\ref{clap}) is junk (since it is a graded commutator), and we are interested in the
exact form of the right hand side. This is easily computed in terms of our established notation, and is
given by
\be A_{jk}+A_{kj}=2f_kf_jId_N.\ee
Taken together, we have shown that
\bea  [\D,a_j][\D,a_k]+[\D,a_k][\D,a_j]=[[\D^2,a_k],a_j]\quad\ \ \nno
\quad =A_{kj}+A_{jk}\ \ \quad\ \ \nno
\quad =2f_kf_jId_N\ \ \quad\ \ \ \nno
\quad\quad\qquad =-2Re(g^i_{jk})Id_N.\eea
This proves that 

1) The $[\D,a^i_j]$ locally generate $Cliff(\Omega^1_\D(a^i_1,...,a^i_p), Re(g^i_{jk}))$, by the universality of the
Clifford relations. Also, from the form of the Hermitian structure on $\Omega^1_\D(\A)$, 
$Re(g^i_{jk})$
is a nondegenerate quadratic form.

2) The operator $\D^2$ is a generalised Laplacian, as $f_kf_j=-Re(g^i_{jk})$.

3) From 2), the principal symbols $\s^{\D^2}_2(x,\xi)=\n\xi\n^2Id$, $\s^{\dd}_1(x,\xi)=\n\xi\n Id$, for
$(x,\xi)\in T^*X|_{U^i}$, the total space of the cotangent bundle over $U^i$. This tells us that $\dd$, $\D^2$ and
$\D$ are elliptic differential operators, at least when restricted to the sets $U^i$. With a very little more
work one can also see that $\s^\D_1(x,\xi)=\xi\cdot$, Clifford multiplication by $\xi$.

4) As $\Omega^*_\D(\A)|_{U^i}\cong{\C}liff(T^*X)|_{U^i}$, and $\HH_\infty$ is an irreducible
module for $\Omega^*_\D(\A)$, we see that $S$ is the (unique) fundamental spinor bundle for $X$; see
\cite[appendix]{ML}.

5) $\D=c\circ\nabla+A$, where $\nabla$ is a compatible connection on the spinor
bundle, and $A$ is a self-adjoint element of $\Omega^1_\D(\A)$. (Using the above results one can
now show that $c\circ\nabla$ is essentially self-adjoint, whence $A$ must be self-adjoint.)

6) It is possible to check that the connection on $\Lambda^*_\D(\A)\otimes\Gamma(X,S)$ given 
by the graded commutator $[\nabla,\cdot]$ is compatible with $(\cdot,\cdot)_{\Omega^1_\D}$. Hence 
$\nabla$ is the lift of a compatible connection on the cotangent bundle. 
\smallskip\newline
The existence of an irreducible representation of ${\C}liff(T^*X\otimes{\cal L})$ for some line 
bundle ${\cal L}$ shows that $X$ is a spin$^c$ manifold. Before completing the proof that $X$ is in fact spin, we briefly
examine the metric. 
\bigskip\newline
It is now some time since 
Connes proved that 
his `sup' definition of the metric coincided with the geodesic distance for the canonical triple on a spin
manifold, \cite{C5}. We will reproduce the proof here for completeness. All one needs to know in order 
to show that
these metrics agree is that for $a\in\A$ the operator 
$[\D,a]=\sum_j(\partial a/\partial a_j)[\D,a_j]$ is (locally, so over $U^i$ for each $i$) 
Clifford
multiplication by the gradient. Then Connes' proof holds with no modification:
\bea\lefteqn{\n[\D,a]\n = \sup_{x\in X} |\sum_{j,k}(\frac{\partial a}{\partial a_j}[\D,a_j])^*
\frac{\partial a}{\partial a_k}[\D,a_k]|^{1/2}}\nno
 &\qquad\, & =\sup_{x\in X}|\sum_{j,k}g_{jk}^i(\frac{\partial a}{\partial a_j})^*
\frac{\partial a}{\partial a_k}|^{1/2}\nno
 &\qquad\, & =\n a\n_{Lip}:=\sup_{x\neq y}\frac{|a(x)-a(y)|}{d_\gamma(x,y)}.\eea
In the last line we have defined the Lipschitz norm, with $d_\gamma(\cdot,\cdot)$ the 
geodesic distance on $X$. 
The constraint $\n[\D,a]\n\leq 1$ forces $|a(x)-a(y)|\leq d_\gamma(x,y)$. To reverse the 
inequality, we fix $x$ and 
observe that $d_\gamma(x,\cdot):X\rightarrow\R$ satisfies $\n[\D,d_\gamma(x,\cdot)]\n\leq 1$. 
Then 
\be \sup\{|a(x)-a(y)|:\ \n[\D,a]\n\leq 1\}=d(x,y)\geq|d_\gamma(x,y)-d_\gamma(x,x)|=
d_\gamma(x,y).\ee
Thus the two metrics $d(\cdot,\cdot)$ and $d_\gamma(\cdot,\cdot)$ agree.

\subsubsection{$X$ is Spin}\label{e}
In discussing the reality condition, we will need to recall that $Cliff_{r,s}$ module multiplication 
is, \cite{ML},

1)  $\R$-linear for $r-s\equiv 0,6,7\bmod 8$

2) $\C$-linear for $r-s\equiv 1,5\bmod 8$

3) $\h$-linear for $r-s\equiv 2,3,4\bmod 8$.
\newline
To show that $X$ 
is spin, we need to show that there exists a representation of $\Omega^*_{\D}(\A_{\R})$, where 
$\A_{\R}=\{a\in\A:a=a^*\}$. This is a real algebra with trivial involution. We will employ the properties of the real structure to do this, also 
extending the treatment to cover representations of $Cliff_{r,s}$, with $r+s=p$. This requires some 
background on Real Clifford algebras, \cite{ML,A}.
\smallskip\newline
Let $Cliff({\R}^{r,s})$ be the Real Clifford algebra on ${\R}^r\oplus{\R}^s$ with positive definite quadratic form 
and involution 
generated by $c:(x_1,...,x_r,y_1,...,y_s)\rightarrow(x_1,...,x_r,-y_1,...,-y_s)$ for $(x,y)\in{\R}^r\oplus{\R}^s$. 
The map $c$ has a unique antilinear 
extension to the complexification ${\C}liff({\R}^{r,s})=Cliff({\R}^{r,s})\otimes\C$ given by $c\otimes cc$, where 
$cc$ is complex conjugation. Note that all the algebras $Cliff_{r,s}$ with $r+s$ the same will become 
isomorphic when complexified, however this is not the case for the algebras $Cliff({\R}^{r,s})$ with the 
involution. If we forget the involution, or if it is trivial, then $Cliff({\R}^{r,s})\cong Cliff_{r+s}$ and 
${\C}liff({\R}^{r,s})\cong{\C}liff_{r+s}$. 
\smallskip\newline
A Real module for $Cliff({\R}^{r,s})$ is a complex representation space for $Cliff_{r,s}$, $W$, along with an 
antilinear map (also called $c$) $c:W\rightarrow W$ such that
\be c(\phi w)=c(\phi)c(w)\ \ \ \forall\phi\in Cliff({\R}^{r,s}),\ \forall w\in W.\ee
It can be shown, \cite{ML}, that the Grothendieck group of Real representations of $Cliff({\R}^{r,s})$ is 
isomorphic to the Grothendieck group of real representations of $Cliff_{r,s}$, and as every Real 
representation of $Cliff({\R}^{r,s})$ automatically extends to ${\C}liff({\R}^{r,s})$, the latter is the 
appropriate complexification of the algebras $Cliff_{r,s}$. It also shows that $KR$-theory is the 
correct cohomological tool.
\smallskip\newline
Pursuing the $KR$ theme a little longer, we note that $(1,1)$-periodicity in this theory corresponds to the 
$(1,1)$-periodicity in the Clifford algebras
\be Cliff_{r,s}\cong Cliff_{r-s,0}\otimes Cliff_{1,1}\otimes\cdot\cdot\cdot\otimes Cliff_{1,1}\ee
where there are $s$ copies of $Cliff_{1,1}$ on the right hand side. As $Cliff_{1,1}\cong M_2(\R)$ is a 
real algebra (as well as Real), this shows why the $\R,\ \C,\ \h$-linearity of the module multiplication 
depends only on $r-s\bmod 8$. We take $Cliff_{1,1}$ to be generated by $1_2$ and $v=(v_1,v_2)\in{\R}^2$ by 
setting
\[ v=\left(\begin{array}{cc}
v_{2} & v_{1} \\
-v_{1} & -v_{2}
\end{array}\right)\]
and the multiplication is just matrix multiplication
\bea v\cdot w=(v_2w_1-v_1w_2)\left(\begin{array}{cc}
 0 & 1 \cr
 1 & 0 \end{array}\right)-(v_1w_1-v_2w_2)1_2\nno
=v\wedge w\left(\begin{array}{cc}0 & -1\\ -1 & 0\end{array}\right)-(v,w)_{1,1}1_2.\eea  
We take $Cliff({\R}^{1,1})$ to be generated by $(v_1,iv_2)$ and we see that the involution is then given by
complex conjugation. The multiplication is matrix multiplication with
\be v\cdot w=-(v,w)_21_2 +v\wedge w\left(\begin{array}{cc}0 & -i\\ -i & 0\end{array}\right).\ee
\smallskip\newline
Thus we may always regard the involution on 
\be {\C}liff({\R}^{r,s})\cong Cliff_{r-s,0}\otimes Cliff({\R}^{1,1})\otimes\cdot\cdot\cdot\otimes Cliff({\R}^{1,1})\otimes\C\ee
as $1\otimes cc\otimes cc\cdot\cdot\cdot\otimes cc\otimes cc$. This is enough of generalities for the moment. In our 
case we have a complex representation space $\Gamma(X,S)$, and an involution $J$ such that 
\be J(\phi\xi)=(J\phi J^*)(J\xi),\ \ \ \forall\phi\in\Omega^*_{\D}(\A),\ \ \forall\xi\in\Gamma(X,S).\ee
So we actually have a representation of ${\C}liff({\R}^{r,s})$ with the involution on the algebra realised by $J\cdot J^*$. It is clear 
that $J\cdot J^*$ has square $1$, and so is an involution, and we set $s=$ number of eigenvalues equal to $-1$. Then from the 
preceeding discussion it is clear that 
\be J\cdot J^*=1_{Cliff_{p-2s,0}}\otimes cc\otimes\cdot\cdot\cdot\otimes cc\otimes cc \ee
with $s$ copies of $cc$ acting on $s$ copies of $Cliff({\R}^{1,1})$ and with the behaviour of $J|_{Cliff_{p-2s,0}}$ 
determined by $p-2s\bmod 8$ according to table (\ref{real}).  It is clear that $J\cdot J^*$ reduces to $1$ on the positive 
definite part of the algebra, as it is an involution with all eigenvalues $1$ there. This implies that $J\cdot J^*$
preserves  elements 
of the form $\phi\otimes 1\otimes\cdot\cdot\cdot\otimes 1\otimes 1_{\C}$ where $\phi\in Cliff_{p-2s,0}$.
However, we still need to fix the behaviour of 
$J$, and this is what is determined by $p-2s\bmod 8$. 
 \smallskip\newline
So we claim that we have a representation of 
$Cliff_{p-s,s}(T^*X,(J\cdot J^*,\cdot)_{\Omega^1})$ provided the behaviour of $J$ is 
determined by $p-2s\bmod 8$ and table \ref{real}.  Two points: First, this reduces to Connes' formulation for $s=0$; second, 
the metric $(J\cdot J^*,\cdot)_{\Omega^1}$ has signature $(p-s,s)$ and making this adjustment corresponds to swapping 
between the multiplication on $Cliff({\R}^{1,1})$ and $Cliff_{1,1}$. Similarly we replace $(\cdot,\cdot)_S$ with $(J\cdot,\cdot)_S$.
\smallskip\newline
In all the above we have assumed that $2s\leq p$. If this is not the case, we may start with the 
negative definite Clifford algebra, $Cliff_{0,2s-p}$, and then tensor on copies of $Cliff_{1,1}$.
\smallskip\newline
Note that it is sufficient to prove the reduction for 
$0<p\leq 8$ and $s=0$. This is 
because the extension to $s\neq 0$ involves tensoring on copies of $Cliff({\R}^{1,1})$ for which the involution is determined, 
whilst raising the dimension simply involves tensoring on a copy of $Cliff_8=M_{16}({\R})$, and this will not affect the following 
argument. These simplifications reduce us to the case $J\cdot J^*=1\otimes cc$ on $\Omega^*_{\D}(\A)$. 
To complete the proof, we proceed by cases.
\smallskip\newline
The first case is $p=6,7,8$. As $J^2=1$ and $J\D=\D J$, $J= cc$. We set $\Gamma_{\R}(X,S)$ to be the fixed point set of $J$. 
Then restricting to the action 
of $\Omega^*_{\D}(\A_{\R})$ on $\Gamma_{\R}(X,S)$, $J$ is trivial. Hence we may regard the representation $\pi$ as 
arising as the complexification of this real representation. As $\phi=J\phi=\phi J=J\phi J^*$ on $\Gamma_{\R}(X,S)$, the 
action can only be $\R$-linear. From the fact that $[\D,J]=0$, we easily deduce that $\nabla
J=0$, so that $J$ is globally parallel. Thus there is no global twisting involved in obtaining
$\Omega^*_\D(\A)$ from $Cliff(T^*X)$. Hence $X$ is spin.
\smallskip\newline
In dimensions $2,3,4$, not only does $J$ commute with $\Omega^*_{\D}(\A_{\R})$, but $i$ does also (we are looking at 
the action on $\Gamma(X,S)$, not $\Gamma_{\R}(X,S)$). So set 
\be e=J,\quad f=i,\quad g=Ji,\ee
note that $e^2=f^2=g^2=-1$, and observe that the following commutation relations hold:
\be ef=-fe=g,\quad fg=-gf=e,\quad ge=-eg=f.\ee
Thus regarding $e,f,g$ and $\Omega^*_{\D}(\A_{\R})$ as elements of $\hom_{\R}(\Gamma(X,S),\Gamma(X,S))$, 
we see that $\Gamma(X,S)$ has the structure of a quaternion vector bundle on $X$, and the action of $Cliff(T^*X)$ is 
quaternion linear. As in the last case, $\nabla J=0$, so that the Clifford bundle is untwisted
and so $X$ is spin.
\smallskip\newline
The last case is $p=1,5$. For $p=1$, the fibres of $\Omega_{\D}^*(\A_{\R})$ are isomorphic to ${\C}$, 
and we naturally have that the Clifford multiplication is ${\C}$-linear. For $p=5$, the fibres are
$M_4({\C})$, and as $J^2=-1$, we have a commuting subalgebra $span_{\R}\{1,J\}\cong{\C}$. Note 
that the reason for the anticommutation of $J$ and $\D$ is that $\D$ maps
real functions to imaginary functions, for $p=1$, and so has a factor of $i$. Analogous statements hold for $p=5$. 
 In particular, removing the complex coefficients, so passing from $\D$ to $\nabla$, we see that $\nabla J=0$, and so
$X$ is spin.       
\smallskip\newline
 Note that in the even dimensional cases when $\pi(c)J=J\pi(c)$, 
         $\pi(c)\in\Omega^{*}_{\D}(\A_{{\mathbf R}}).$ When they anticommute, 
$\pi(c)$ is $i$ times a real form. 
         This corresponds to the behaviour of the complex volume form of a spin 
         manifold on the spinor bundle. Compare the above discussion 
         with \cite{ML}. 
\smallskip\newline
It is interesting to consider whether we can recover the indefinite distance from $(J\cdot J^*,\cdot)_{\Omega^1}$. We 
will not address the issue here, but simply point out that in the topology determined by $(J\cdot J^*,\cdot)$, 
our previously compact space is no longer 
necessarily compact, and so can not agree with the weak$^*$ topology. 
It is worth noting that if $J\cdot J^*$ has one or more negative eigenvalues and $\nabla$ is compatible with 
the Hermitian form $(J\cdot,\cdot)_S$, then $\D=c\circ\nabla$ is hyperbolic rather than elliptic. So many remaining 
points of the proof, relying on the ellipticity of $\D$, will not go through for the pseudo-Riemannian case. We will however 
point out the occasional interesting detail for this case. 
     \smallskip\newline
      So for all dimensions we have shown that $X$ is a spin manifold with $\A$ the smooth functions on 
$X$ acting as multiplication 
operators on an irreducible spinor bundle. Thus $3)$ is proved completely.

\subsection{Completion of the Proof.}
\subsubsection{Generalities and Proof of $4)$.}
To prove $4)$, note that if we make a unitary change of representation, the 
metric, the integration defined via the Dixmier trace, and the absolutely continuous spectrum 
of the $a^i_j$ (i.e. $X$), are all unchanged. 
The only object in sight that varies in any important way with unitary change of representation 
is the operator $\D$. The 
change of representation induces an affine change on $\D$:
\be \D\rightarrow U\D U^*=\D+U[\D,U^*].\ee
This in itself shows that the connected components of the fibre over 
$[\pi]\rightarrow d_\pi(\cdot,\cdot)$ are affine. 
To show that there are a finite number of components, it suffices to note 
that a representation in any component 
satisfies the axioms, (recall that a spin structure for one metric canonically determines one for 
any other metric, \cite{ML}), and so gives rise to an action of the Clifford bundle, and 
so to a spin structure. As there are only 
a finite number of these, we have proved $4)$.
\smallskip\newline
The only items remaining to be proved are, for $p>2$, 

1. $\intcross\dd^{2-p}$ is a positive definite quadratic form on each $A_\sigma$ with 
unique minimum $\pi_\sigma$

2. This minimum is achieved for $\D=\Dslash$, the Dirac operator on $S_\sigma$

3. $\intcross|\Dslash|^{2-p}=-\frac{(p-2)c(p)}{12}\int_XRdv$.
\newline
These last few items will all be proved by direct computation once 
we have narrowed down the nature of $\D$ a bit 
more. As an extra bonus, we will also be able to determine the measure once we 
have this extra information.
\smallskip\newline
Recall the condition for compatibility of a 
connection $\nabla^S$ on $S$ with the Hermitian structure $(\cdot,\cdot)_S$ as 
\be [\D,(\xi,\eta)_S]= (\xi,\nabla^S\eta)_S-(\nabla^S\xi,\eta)_S,\ \ 
\forall\xi,\eta\in\Gamma(X,S).\ee 
 Given such a connection, the graded commutator $[\nabla^S,\cdot]:\Lambda^*_{\D}(\A_{\R})\rightarrow
\Lambda^{*+1}_{\D}(\A_{\R})$ is a connection 
compatible 
with the metric on $\Lambda^*_{\D}(\A_{\R})$. If instead we have a connection compatible with $(J\cdot,\cdot)_S$, then
$[\nabla^S,\cdot]$ is compatible with $(J\cdot J^*,\cdot)_{\Omega^1_\D}$. Note that we are really considering differential 
forms with values in $\Gamma(X,S)$, so actually have a connection  
$[\nabla^S,\cdot]:\Lambda^*_{\D}(\A)\otimes\Gamma(X,S)\rightarrow\Lambda^{*+1}_{\D}(\A)
\otimes\Gamma(X,S).$ Beware of confusing the notation here, for $[\nabla^S,\cdot]$ uses the graded commutator,
while $[\D,\cdot](a[\D,b])=[\D,a][\D,b]$.  
\smallskip\newline
The torsion of the connection $[\nabla^S,\cdot]$ on $T^*X$  
is defined to be $T([\nabla^S,\cdot])=d-\epsilon\circ[\nabla^S,\cdot]$, where 
$d=[\D,\cdot]$ and $\epsilon$ is just antisymmetrisation. 
Then from what has been proved thus far, we have
\be \D=c\circ\nabla^S+ T,\ \ [\D,\cdot]=c\circ[\nabla^S,\cdot]+ c\circ T([\nabla^S,\cdot]),\ee
on $\Gamma(X,S)$ and $\Omega^*_{\D}(\A_{\R})\otimes\Gamma(X,S)$ respectively. Here $c$ is the 
composition of Clifford multiplication with the derivation in question. On the bundle
$\Lambda^*_\D(\A)\otimes\Gamma(X,S)$ we have already seen that $[\D,\cdot]$ is the exterior derivative. 
The $T$ in the expression for $\D$ is the lift of the
torsion term to the spinor bundle. 
\smallskip\newline
Any two compatible connections on $S$ differ by a $1-$form, $A$ say, and by virtue of the 
first order condition, 
adding $A$ to $\nabla^S$ does not affect $[\nabla^S,\cdot]$, and so in particular $\nabla^S$ would still be the 
lift of a compatible connection on the cotangent bundle. As $U[\D,U^*]$ is self-adjoint, for any representation $\pi$, the 
operator $D_\pi$ is the Dirac operator of a compatible connection on the spinor bundle. Note that as $\D$ is self-adjoint, the Clifford action
of any such 1-form $A$ must be self-adjoint on the spinor bundle. 
\smallskip\newline
It is important to note that for every unitary element of the algebra, $u$ 
say, gives rise to a unitary transformation $U=uJuJ^*$. If we start with $\D+A$, $A=A^*\in\Omega^1_{\D}(\A)$, 
and conjugate by 
$U$, we obtain $\D+A+JAJ^*+u[\D,u^*]+Ju[\D,u^*]J^*$. If the metric is positive definite, then 
$JAJ^*=-A^*$ for all $A\in\Omega^*_{\D}(\A)$. Thus all of these gauge terms (or internal 
fluctuations, \cite{C1}) vanish in the positive definite, commutative case. This corresponds to the Clifford algebra being
built on the untwisted cotangent bundle, so that we do not have any $U(1)$ gauge terms. In the indefinite case we 
find non-trivial gauge terms associated with timelike directions. To see this, note that every element of 
$\Omega^1_{\D}(\A)$ is of the form $A+iB$, where each of $A$ and $B$ are real, so anti-self-adjoint. Possible 
gauge terms are of the form $iB$, as they must be self-adjoint. If we assume that $B$ is timelike (i.e. $JBJ^*=-B$), 
and set $(u[\D,u^*])_t$ to be the timelike part of $u[\D,u^*]$, then 
\bea\lefteqn{U(\D+iB)U^*=\D+iB+JiBJ^*+u[\D,u^*]+Ju[\D,u^*]J^*}\nno
&\qquad &\qquad\ \, =\D+iB-iJBJ^*+u[\D,u^*]+Ju[\D,u^*]J^*\nno
&\qquad &\qquad\ \, =\D+2iB+2(u[\D,u^*])_t.\eea
Thus we can find non-trivial gauge terms in timelike directions. 
\bigskip\newline
Since we are unequivocably in the manifold setting now, and as we shall require the symbol 
calculus to 
compute the Wodzicki residue, we shall now change notation. In traditional fashion, let us write
\be \gamma^\mu\gamma^\nu+\gamma^\nu\gamma^\mu=-2g^{\mu\nu}1_S\ee
\be \gamma^a\gamma^b+\gamma^b\gamma^a=-2\delta^{ab}1_S\ee
for the curved (coordinate) and flat (orthonormal) gamma matrices respectively. Let $\sigma^k$, 
$k=1,...,[p/2]$, be a local 
orthonormal basis of $\Gamma(X,S)$, and $a\in\pi(\A)$. Then the most general form that $\D_\pi$ can take is
\be \D_\pi(a\sigma^k)=\sum_\mu\gamma^\mu(\partial_\mu a)\sigma^k +
\frac{1}{2}\sum_{\mu,a<b}a\gamma^\mu\omega_{\mu ab}\gamma^a\gamma^b\sigma^k + 
\frac{1}{2}\sum_{\mu,a<b}a\gamma^\mu t_{\mu ab}\gamma^a\gamma^b\sigma^k +a\sum_\mu\gamma^\mu f_\mu\s^k\ee
where $\omega$ is the lift of the Levi-Civita connection to the bundle of spinors,  
$t$ is the lift of the torsion term, and $f_\mu$ is a gauge term associated to timelike directions. We assume without loss 
of generality that our coordinates allow us to split the cotangent space so that timelike and spacelike terms are 
orthogonal. Then we may take $f_\mu=0$ for $\mu$ the index of a spacelike direction. 
We will now drop the $\pi$ and consider $\D$ as being determined by 
$t$ and $f_\mu$. It is worth noting that $t_{cab}$ is totally antisymmetric, where $t_{\mu ab}=e^c_\mu t_{cab}$, and 
$e^c_\mu$ is the vielbein. Note that from our previous discussion, the appropriate choice of Dirac operator is no longer elliptic, 
and so in the following arguments we will assume that $J\cdot J^*$ has no negative eigenvalues. Thus from this point on
we assume that we are in the positive definite case with $f_\mu=0$ and $\D=\D(t)$. 
\smallskip\newline
This gives us enough information to recover the measure on our space also. All of these operators, $\D(t)$, have the same 
principal symbol, $\xi\cdot$, Clifford multiplication by $\xi$. Hence, over the unit sphere bundle the principal symbol 
of $\dd$ is $1$. Likewise, the restriction of the principal symbol of $a\dd^{-p}$ to the unit sphere bundle is $a$, where 
here we mean $\pi(a)$, of course. Before evaluating the Dixmier trace of $a\dd^{-p}$, let us look at the 
volume form.
\smallskip\newline
Since the $[\D,a^i_j]$ are independent at each point of $U^i$, the sections 
$[\D,a^i_j]$, $j=1,...,p$, form a (coordinate) basis of the cotangent bundle. Then their product is the real 
volume form $\omega^i$. With $\omega_{\C}=i^{[(p+1)/2]}\omega$ the complex volume form, we have
\be\Gamma=\pi(c)=\sum_ia^i_0[\D,a^i_1]\cdot\cdot\cdot[\D,a^i_p] =\sum_i\tilde{a}^i_0\omega_{\C}^i\ee
where $a^i_0=\tilde{a}^i_0i^{[(p+1)/2]}$, \cite{ML}. 
\smallskip\newline
As $\omega_{\C}$ is central over $U^i$ for $p$ odd, it must be a scalar multiple, $k$, of the identity. So 
$\sum_ik\tilde{a}^i_0(x)=k$, and we see that the collection of maps $\{\tilde{a}^i_0\}_i$ form a partition of unity 
subordinate to the $U^i$. The axioms tell us that $k=1$. In the even case, $\omega_{\C}$ gives the 
${\bf Z}_2$-grading of the Hilbert space,
\be \HH=\frac{1+\omega_{\C}}{2}\HH\oplus\frac{1-\omega_{\C}}{2}\HH.\ee
This corresponds to the splitting of the spin bundle, and for sections of these subbundles we have
\be 1=\sum_i\tilde{a}^i_0\frac{1+\omega^i_{\C}}{2}=\sum_i\tilde{a}^i_0\ee
and similarly for $\frac{1-\omega_{\C}}{2}$. Thus in the even dimensional case we also have a partition of unity.
\smallskip\newline
Recall the usual definition of the measure on $X$. To integrate a function $f\in\A$ over a single coordinate chart 
$U^i$, we make use of the (local) embedding $a^i:U^i\rightarrow {\R}^p$. We write
$f=\tilde{f}(a^i_1,...,a^i_p)$ where $\tilde{f}:{\R}^p\to{\C}$ has compact support. Then
\be \int_{U^i}f:=\int_{U^i}(a^i)^*(\tilde{f})=\int_{a^i(U^i)}\tilde{f}(x)d^px.\ee
To integrate $f$ over $X$, we make use of the embedding $a$ and the partition of unity and write
\be \sum_i\int_{a^i(U^i)}(\tilde{a}^i_0\tilde{f})(x)d^px.\ee
\smallskip\newline
Now given a smooth space like $X$, a 
representation of the continuous functions will split into two pieces; one absolutely continuous with respect to the 
Lebesgue measure, above, and one singular with respect to it, \cite{V}, $\pi=\pi_{ac}\oplus\pi_s$. This gives us a 
decomposition of the Hilbert space 
into complementary closed subspaces, $\HH=\HH_{ac}\oplus\HH_{s}$. The joint spectral measure of the $a^i_j,$ 
$j=1,...,p$, is absolutely continuous with respect to the $p$ dimensional Lebesgue measure, so 
$\overline{\HH_\infty}\subseteq\HH_{ac}$. By the definition of the inner product on $\HH_\infty$ given in 
the axiom of finiteness and absolute continuity, $\overline{\HH_\infty}=L^2(X,S,\intcross\cdot\dd^{-p})$. As the 
Lebesgue measure on the 
joint absolutely continuous spectrum is itself absolutely continuous with respect to the measure given by the 
Dixmier trace, we must also have $\HH_{ac}\subseteq\overline{\HH_\infty}$, and so they are equal. As all the 
$a^i_j$ act as zero on $\HH_s$, recall they are smooth elements, and they generate both $\A$ and $\overline{\A}$, 
the requirement of irreducibilty says that $\HH_s=0$. Thus the representation is absolutely continuous, and as the 
measure is in the same measure class as the Lebesgue measure, $\HH=\overline{\HH_\infty}=L^2(X,S)$.
\smallskip\newline
Let us now compute the value of the integral given by the Dixmier trace. From the form of $\D$,
we know that $\D$ is an operator of order $1$ on the spinor bundle of $X$, so $\dd^{-p}$ is of order $-p$. 
Invoking Connes' trace theorem

 \bea\lefteqn{\bigintcross 
         f\dd^{-p} 
         =\frac{1}{p(2\pi)^{p}}\sum_i\int_{S^{*}U^i}tr_{S}(\tilde{a}^i_0f)\sqrt{g}d^{p}xd\xi}\nno
&\qquad &\quad =\frac{2^{[p/2]}Vol(S^{p-1})}{p(2\pi)^{p}}\sum_i\int_{U^i}\tilde{a}^i_0f\sqrt{g}d^{p}x.\eea
         Thus the inner product on $\HH$ is given by
         \be \langle 
         a\xi,\eta\rangle= 
\frac{2^{[p/2]}Vol(S^{p-1})}{p(2\pi)^{p}}\int_X(a^*(\xi,\eta)_S\sqrt{g})(x)d^{p}x.\ee
         We note for future reference that $\mbox{Vol}(S^{p-1})= 
         \frac{(4\pi)^{p/2}}{2^{p-1}\Gamma(p/2)}$, \cite{G}, so that the complete 
factor above is the same as in equation (\ref{const}), 
               \be           \frac{2^{[p/2]}Vol(S^{p-1})}{p(2\pi)^{p}}=c(p).\ee
All the above discussion is limited to the case $p\neq 1$. The only $1$ dimensional compact spin manifold is $S^1$. In 
this case the Dirac operator is $\frac{1}{i}\frac{d}{dx}$, with singular values $\mu_n(\dd^{-1})=\frac{1}{n}$. In 
\cite[pp 311,312]{C}, Connes presents an argument bounding the $(p,1)$ norm of $[f(\epsilon\D),a]$ in terms of 
$[\D,a]$ and the Dixmier trace of $\D$, with $\epsilon>0$ and $f$ a smooth, even, compactly supported, real function. 
From this 
Theorem \ref{bound} is a consequence of specialising $f$. 
Our aim then is to bound the trace of $[f(\epsilon\D),a]$. So suppose 
that the support of $f$ is contained in $[-k,k]$. Then the rank of $[f(\epsilon\D),a]$ is bounded by the number of 
eigenvalues of $\dd^{-1}$ $\geq \epsilon k^{-1}$. Calling this number $N$, we have $N\leq\epsilon^{-1}k$ and so 
\be \n [f(\epsilon\D),a]\n_1\leq 2\epsilon^{-1}k\n[f(\epsilon\D),a]\n.\ee
The rest of the argument uses Fourier analysis techniques to bound the commutator in terms of $\n[\D,a]\n$, and noting that 
$\sum^N\frac{1}{n}/\log N\geq 1$, for then
\be   \n [f(\epsilon\D),a]\n_1\leq 2C_f\n [\D,a]\n\bigintcross\dd^{-1}.\ee
From this a choice of $f$ gives the analogue of Theorem \ref{bound} in this case, as the results of 
Voiculescu and Wodzicki hold for dimension $1$; see \cite{C} for the full story.
\smallskip\newline
As the Dirac operator of a compatible Clifford connection is self-adjoint only when there is no boundary, the
self-adjointness of $\D$ and the geometric interpretation of the inner product on the Hilbert space now shows that the
spin manifold $X$ is closed. There are numerous consequences of closedness, as well as a more general formulation for
the noncommutative case; see \cite{C}. All that remains is to examine the gravity action given by the
Wodzicki residue.

\subsubsection{The Even Dimensional Case.}
Much of what follows is based on \cite{KW}, though we also complete the odd dimensional case. We also 
note that this calculation was carried out in the four dimensional case in \cite{DK}.
 The key to the 
following computations is the composition formula for symbols:
          \be \sigma(P\circ Q)(x,\xi)=\sum^{\infty}_{|\alpha|=0}\frac{(-i)^{|\alpha|}}{\alpha !}
          (\partial^{\alpha}_{\xi}\sigma(P))(\partial^{\alpha}_{x}\sigma(Q)).\ee
We shall use this to determine $\s_{-p}(\dd^{2-p})$, so that we may compute the Wodzicki residue. 
In the even dimensional case, we use this formula to obtain the following,
  \bea\lefteqn{\sigma_{-p}(\D^{2-p})=\sigma_{0}(\D^{2})\sigma_{-p}(\D^{-p})+\sigma_{1}(\D^{2})
          \sigma_{-p-1}(\D^{-p})}\nno
&\qquad &\quad\,  +\sigma_{2}(\D^{2})\sigma_{-p-2}(\D^{-p}) -i\sum_{\mu}
          (\partial_{\xi_{\mu}}\sigma_{1}(\D^{2}))(\partial_{x^{\mu}}\sigma_{-p}(\D^{-p}))\nno
&\qquad &\quad\, -i\sum_{\mu}(\partial_{\xi_{\mu}}\sigma_{2}(\D^{2}))(\partial_{x^{\mu}}
\sigma_{-p-1}(\D^{-p}))\nno
&\qquad &\quad\,-\frac{1}{2}\sum_{\mu,\nu}(\partial^{2}_{\xi_{\mu}\xi_{\nu}}\sigma_{2}(\D^{2}))
          (\partial^{2}_{x^{\mu}x^{\nu}}\sigma_{-p}(\D^{-p})).\eea
This involves the symbol of $\D^2$ which we can compute, and lower order terms from
$\dd^{-p}$. Since $\dd^2=\D^2$, we have a simplification in the even dimensional case, namely that the expansion
  \be \sigma(\D^{-2m})=\sum_{|\alpha|=0}^{\infty}\frac{(-i)^{|\alpha|}}{\alpha !}
(\partial^{\alpha}_{\xi}\sigma(\D^{-2m+2})(\partial^{\alpha}_{x}\sigma(\D^{-2})),\label{even}\ee
provides a recursion relation for the lower order terms provided we can determine the first few
terms of the symbol for a parametrix of $\D^2$. Let $\s_2=\s_2(\D^2)$ and $p=2m$. 
Then by the
multiplicativity of principal symbols, or from the above, $\s_{-2m}(\D^{-2m})=\s_2^{-m}$, at least away from the zero
section. Also
let us briefly recall that while principal symbols are coordinate independent, other terms are not. So
all the following calculations will be made in Riemann normal coordinates, for which the metric
takes the simplifying form
  \be g^{\mu\nu}(x)=\delta^{\mu\nu}-\frac{1}{3}R^{\mu\nu}_{\rho\sigma}(x_{0})
          x^{\rho}x^{\sigma}+O(x^{3}).\ee
This choice will simplify many expressions, and we will write $=_{RN}$ to denote equality in these 
coordinates. Also, as we will be interested in the value of certain expressions on the cosphere 
bundle, we will also employ the symbol $=_{RN,\bmod\n\xi\n}$ to denote a Riemann normal 
expression in which $\n\xi\n$ has been set to $1$.
So using (\ref{even}) to write
\bea\lefteqn{\sigma_{-2m-1}(\D^{-2m})=\sigma_{2}^{-m+1}\sigma_{-3}(\D^{-2})+\sigma_{-2m+1}(\D^{-2m+2})
\sigma_{2}^{-1}}\nno
&\qquad &\qquad\quad -i\sum_{\mu}(\partial_{\xi_{\mu}}
\sigma_{2}^{-m+1})
          (\partial_{x^{\mu}}\sigma_{2}^{-1}),\eea
we can use Riemann normal coordinates to simplify this to
 \bea\lefteqn{\sigma_{-2m-1}(\D^{-2m})=_{RN}\sigma_{2}^{-m+1}\sigma_{-3}(\D^{-2})}\nno
&\qquad\qquad\quad\ \,+\sigma_{-2m+1}(\D^{-2m+2})\sigma_{2}^{-1}\nno
&\qquad\qquad\ \quad\,\,\,=_{RN}m\sigma_{2}^{-m+1}\sigma_{-3}(\D^{-2}),\eea
          after applying recursion in the obvious way. The next term to compute is 

          \bea\lefteqn{\sigma_{-2m-2}(\D^{-2m})=\sigma^{-m+1}_{2}\sigma_{-4}(\D^{-2})+\sigma_{-2m+1}(\D^{-2m+2})
\sigma_{-3}(\D^{-2})}\nno
&\qquad &\qquad\quad  +\sigma_{-2m}(\D^{-2m+2})\sigma_{2}^{-1}
         -i\sum_{\mu}(\partial_{\xi_{\mu}}\sigma_{2}^{-m+2})(\partial_{x^{\mu}}
\sigma_{-3}(\D^{-2}))\nno
&\qquad &\qquad\quad -\frac{1}{2}\sum_{\mu,\nu}(\partial^{2}_{\xi_{\mu}\xi_{\nu}}\sigma_{2}^{-m+1})
          (\partial^{2}_{x^{\mu}x^{\nu}}\sigma_{2}^{-1}).\eea
Using the last result and the following two expressions

          \[\partial^{2}_{\xi_{\mu}\xi_{\nu}}\parallel\xi\parallel^{-2m+2}=_{RN}2m(2m-2)
          \sigma_{2}^{-m-1}\delta^{\mu\tau}\xi_{\tau}\delta^{\nu\sigma}\xi_{\sigma}
          -(m-1)\sigma_{2}^{-m}\delta^{\mu\nu},\]

          \[ \partial^{2}_{x^{\mu}x^{\nu}}\parallel\xi\parallel^{-2}=_{RN}
          \frac{1}{3}R^{\rho\sigma}_{\mu\nu}\xi_{\rho}\xi_{\sigma}\sigma_{2}^{-2}\]
we find

          \bea\lefteqn{\s_{-2m-2}(\D^{-2m})=_{RN}\s_{2}^{-m+1}\s_{-4}(\D^{-2})+(m-1)\s_{2}^{-m+2}(\s_{-3}(\D^{-2}))^{2}}\nno
&\qquad &\qquad\quad +\s_{-2m}(\D^{-2m+2})\s_{2}^{-1}+2i(m-1)\delta^{\mu\sigma}\xi_{\s}\partial_{x^{\mu}}\s_{-3}(\D^{-2})\nno
&\qquad &\qquad\quad -\frac{4m(m-1)}{3}\s_{2}^{-m-3}\xi^{\mu}\xi^{\nu}R^{\rho\s}_{\mu\nu}\xi_{\rho}\xi_{\s}
          +\frac{(m-1)}{3}\s_{2}^{-m-2}\delta^{\mu\nu}R^{\rho\s}_{\mu\nu}\xi_{rho}\xi_{\s}.\eea
Given $\s_{-4}(\D^{-2})$ and $\s_{-3}(\D^{-2})$ this can be computed recursively, giving 

          \bea\lefteqn{\s_{-2m-2}(\D^{-2m})=_{RN}m\s_{2}^{-m+1}\s_{-4}(\D^{-2})+
          \frac{m(m-1)}{2}\s_{2}^{-m+2}(\s_{-3}(\D^{-2}))^{2}}\nno
&\qquad &\qquad\quad +im(m-1)\xi^{\mu}\partial_{x^{\mu}}\s_{-3}(\D^{-2})\nno
&\qquad &\qquad\quad -\frac{4m(m+1)(m-1)}{9}\s_{2}^{-m-3}\xi^{\mu}\xi^{\nu}R^{\rho\s}_{\mu\nu}
\xi_{\rho}\xi_{\s}\nno
&\qquad &\qquad\quad +\frac{m(m-1)}{6}\s_{2}^{-m-2}\delta^{\mu\nu}R^{\rho\s}_{\mu\nu}\xi_{\rho}\xi_{\s},\eea 
where $\xi^{\mu}=\delta^{\mu\nu}\xi_{\nu}.$ In the even case, this gives us a short cut; 
we shall 
          compute this term in general for the odd case, but note that in the even case the short cut 
          gives us

          \bea\lefteqn{\s_{-p}(\D^{-p+2})=_{RN,\bmod\parallel\xi\parallel}\frac{(p-2)}{2}\s_{-4}(\D^{-2}) 
  +\frac{(p-2)(p-4)}{8}(\s_{-3}(\D^{-2}))^{2}}\nno
&\qquad &\qquad +\frac{(p-2)(p-4)}{4}i\xi^{\mu}\partial_{\mu}\s_{-3}(\D^{-2})
          -\frac{p(p-2)(p-4)}{18}\xi^{\mu}\xi^{\nu}R^{\rho\s}_{\mu\nu}\xi_{\rho}\xi_{\s}\nno
&\qquad &\qquad  +\frac{(p-2)(p-4)}{24}\delta^{\mu\nu}R^{\rho\s}_{\mu\nu}\xi_{\rho}\xi_{\s}.\eea
       Having obtained $\s_{-2m-2}(\D^{-2})$ and $\s_{-2m-1}(\D^{-2})$, the next step is to 
          compute $\s_{-3}(\D^{-2})$ and $\s_{-4}(\D^{-2})$. We 
          follow the method of \cite{KW} to construct a 
          parametrix for $\D^{2}$. 
\smallskip\newline
First, let us write $\D^2$ in elliptic operator form
  \be \D^{2}=-g^{\mu\nu}\partial_{\mu}\partial_{\nu}+a^{\mu}\partial_{\mu}+b.\ee
          So the symbol of $\D^{2}$ is 
          \bea\lefteqn{\sigma(\D^{2})=g^{\mu\nu}\xi_{\mu}\xi_{\nu}+ia^{\mu}\xi_{\mu}+b}\nno
&\qquad   =\parallel\xi\parallel^{2}+ia^{\mu}\xi_{\mu}+b\nno
&\    =\sigma_{2}+\sigma_{1}+\sigma_{0}.\eea
          With this notation in hand, let $P$ be the pseudodifferential operator defined 
          by $\s(P)=\s_{2}^{-1}$. In fact we should consider the product $\chi(|\xi|)\s_2(x,\xi)^{-1}$, where 
$\chi$ is a smooth function vanishing for small values of its (positive) argument. As this does not affect the following argument, 
only altering the result by an infinitely smoothing operator, we shall omit further mention of this ''mollifying function''.  
So, one readily checks that $\s(\D^{2}P-1)$ is a symbol of 
          order $-1$. Denoting this symbol by $r$, we have 
          \be \s(\D^{2}P)=1+r \ \ \ \mbox{so}\ \ \s(\D^{2}P)\circ(1+r)^{-1}\sim 1\ee 
          where on the right composition means the symbol of the composition of operators. So 
          if $\s(R)=1+r$, then $\D^{2}PR^{-1}\sim 1$. Hence $PR^{-1}\sim \D^{-2}$. As $r$ is of order $-1$, we 
          may expand $(1+r)^{-1}$ as a geometric series in symbol space. Thus
          \bea\lefteqn{\s(\D^{-2})\ \ \sim \s_{2}^{-1}\circ\sum_{k=0}^{\infty}(-1)^{k}r^{\circ k}}\nno
&\!\sim \s_{2}^{-1}\circ(1-r+r^{\circ 2}-r^{\circ 3}+\ldots)\nno
&\quad\  \ \  \ \,\ \ \sim \s_{2}^{-1}-\s_{2}^{-1}\circ r +\s_{2}^{-1}\circ r\circ r +\mbox{order -5}.\eea
          It is straightforward to compute the part of order $-1$ of $r$

          \bea\lefteqn{r_{-1}=ia^{\mu}\xi_{\mu}\s_{2}^{-1}+
          2i\xi^{\mu}g^{\rho\tau}_{,\mu}\xi_{\rho}\xi_{\tau}\s_{2}^{-1}}\nno
&\  \  \,=_{RN}ia^{\mu}\xi_{\mu}\s_{2}^{-1}\eea
 and its derivative

          \be \partial_{x^{\mu}}r_{-1}=_{RN}ia^{\rho}_{,\mu}\xi_{\rho}\s_{2}^{-1}
          -\frac{2i}{3}\xi^{\rho}R^{\alpha\tau}_{\rho\mu}\xi_{\alpha}\xi_{\tau}\s_{2}^{-2},\ee
 as well as the part of order $-2$

          \be r_{-2}=_{RN}b\s_{2}^{-1}-
\frac{2}{3}\delta^{\mu\nu}R^{\rho\s}_{\mu\nu}\xi_{\rho}\xi_{\s}\s_{2}^{-2}.\ee
Using the composition formula (repeatedly) and discarding terms of order $-5$ or less, we 
          eventually find that

          \be \s_{-3}(\D^{-2})=-ia^{\mu}\xi_{\mu}\s_{2}^{-2},\ee
 and

          \bea\lefteqn{\s_{-4}(\D^{-2})=-b\s_{2}^{-2}+
          \frac{2}{3}\delta^{\mu\nu}R^{\alpha\tau}_{\mu\nu}\xi_{\alpha}\xi_{\tau}\s_{2}^{-3}}\nno
&\qquad\quad\ \,+2\xi^{\mu}a^{\rho}_{,\mu}\xi_{\rho}\s_{2}^{-3}-
a^{\mu}\xi_{\mu}a^{\rho}\xi_{\rho}\s_{2}^{-3}\nno
&\ \!-\frac{4}{3}\xi^{\mu}\xi^{\nu}R^{\alpha\tau}_{\nu\mu}\xi_{\alpha}\xi_{\tau}\s_{2}^{-4}.\eea
           Employing the shortcut for the even case yields 

          \bea\lefteqn{\s_{-p}(\D^{-p+2})=_{RN,\bmod\parallel\xi\parallel}
          -\frac{1}{2}(p-2)b+\frac{p(p-2)}{4}\xi^{\mu}a^{\rho}_{,\mu}\xi_{\rho}}\nno
&\qquad &\qquad -\frac{p(p-2)}{8}a^{\mu}\xi_{\mu}a^{\rho}\xi_{\rho}+
          \frac{p(p-2)}{24}\delta^{\mu\nu}R^{\rho\s}_{\mu\nu}\xi_{\rho}\xi_{\s}\nno
&\qquad &\qquad -\frac{(p-2)(p^{2}-4p+6)}{18}\xi^{\mu}\xi^{\nu}R^{\rho\s}_{\mu\nu}\xi_{\rho}\xi_{\s}.\eea
           In order to perform the integral over the cosphere bundle, we make use of the standard results

          \be\int_{\parallel\xi\parallel=1}\xi^{\mu}d\xi=0,\ \ 
          \int_{\parallel\xi\parallel=1}\xi^{\mu}\xi^{\nu}\xi^{\rho}d\xi=0,\ \ 
\int_{\parallel\xi\parallel=1}\xi^{\mu}\xi^{\nu}d\xi=\frac{1}{p}g^{\mu\nu},\ee
and

          \be\int_{\parallel\xi\parallel=1}\xi^{\mu}\xi^{\nu}\xi^{\rho}\xi^{\s}d\xi=\frac{1}{p(p+2)}
          (g^{\mu\nu}g^{\rho\s}+g^{\mu\rho}g^{\nu\s}+g^{\s\nu}g^{\mu\rho}).\ee
Using the symmetries of the Riemann tensor, one may use the last result to show that

          \be\int_{S^{*}X}R^{\mu\nu}_{\alpha\beta}\xi^{\alpha}\xi^{\beta}\xi^{\s}\xi^{\tau}
          g_{\s\mu}g_{\tau\nu}d\xi dx=0.\ee
Thus 
          \bea\lefteqn{WRes(\D^{2-p})=\frac{1}{p(2\pi)^{p}}\int_{S^{*}X}tr\s_{-p}(\D^{2-p})\sqrt{g}d\xi dx}\nno
&\qquad &\qquad\, =-\frac{(p-2)}{2}\frac{Vol(S^{p-1})}{p(2\pi)^{p}}\int_{X}tr(b+\frac{1}{4}a^{\mu}a_{\mu}
-\frac{1}{2}a^{\mu}_{,\mu})\sqrt{g}dx\nno
&\qquad &\qquad\,+\frac{(p-2)}{24}\frac{Vol(S^{p-1})2^{[p/2]}}{p(2\pi)^{p}}\int_{X}R\sqrt{g}dx.\eea
      \smallskip\newline
To make use of this we will need expressions for $a^\mu$ and $b$. The art of squaring Dirac operators is well 
described in the literature, and we follow \cite{KW}. Writing 
\be \D=\gamma^\mu(\nabla_\mu +T_\mu )\ee
the square may be written, with $\nabla$ the lift of the Levi-Civita connection,
\be \D^2=-g^{\mu\nu}(\nabla_\mu\nabla_\nu)+(\Gamma^\nu-4T^\nu)(\nabla_\nu+T_\nu)
+\frac{1}{2}\gamma^\mu\gamma^\nu[\nabla_\mu+T_\mu,\nabla_\nu+T_\nu].\ee
Here we have used the formulae

$\gamma^\mu[T_\mu,\gamma^\nu]=-4T^\nu$

$\gamma^\mu[\nabla_\mu,\gamma^\nu]=-\gamma^\mu\gamma^\rho\Gamma^\nu_{\mu\rho}=\Gamma^\nu:=
g^{\mu\rho}\Gamma^\nu_{\mu\rho}$
\smallskip\newline
To simplify the following, we also make use of the fact that the Christoffel symbols and their partial derivatives vanish in 
Riemann normal coordinates, and $\gamma^{\mu\nu}[\nabla_\mu,\nabla_\nu]=\frac{1}{2}R$, with $R$ the scalar curvature. 
We can then read off 
\be a^\mu=-2(\omega^\mu+3T^\mu)\ee
\bea\lefteqn{b= \frac{1}{2}a^\mu_{,\mu}-\frac{1}{4}a^\mu a_\mu +5T^\mu_{,\mu}+2[\omega^\mu,T_\mu]}\nno
 &\ \  +4T^\mu T_\mu
+\frac{1}{4}R+\gamma^{\mu\nu}[\nabla_\mu,T_\nu]+\frac{1}{2}\gamma^{\mu\nu}[T_\mu,T_\nu].\eea
As $[\omega^\mu,T_\mu]=g^{\mu\nu}[\omega_\nu,T_\mu]=0$, and the trace of $T^\mu_{,\mu}$ and 
 vanishes, we have
\be trace_S(b+\frac{1}{4}a^\mu a_\mu -\frac{1}{2}a^\mu_{,\mu})=2^{[p/2]}(\frac{1}{4}R-3t_{abc}t^{abc})+
trace_S(\gamma^{\mu\nu}[\nabla_\mu,T_\nu]).\ee
Here we have used

$trace_S(T^\mu T_\mu)=-\frac{1}{2}t_{abc}t^{abc}\times 2^{[p/2]}$

$trace_S(\frac{1}{2}\gamma^{\mu\nu}[T_\mu,T_\nu])=-t_{abc}t^{abc}\times 2^{[p/2]}.$
\smallskip\newline
 So for the even case we arrive at

          \bea\lefteqn{WRes(\D^{-p+2})=-\frac{(p-2)Vol(S^{p-1})2^{[p/2]}}{12p(2\pi)^{p}}\int_{X}R\sqrt{g}dx}\nno
&\qquad &\qquad\ \  -\frac{(p-2)Vol(S^{p-1})}{2p(2\pi)^{p}}
\int_{X}(-3t_{abc}t^{abc}+tr(\gamma^{\mu\nu}[\nabla_{\mu},T_{\nu}]))\sqrt{g}dx.\eea
          As $\nabla_{\mu}$ is torsion free, $\gamma^{\mu\nu}[\nabla_{\mu},T_{\nu}]$ is a boundary 
          term, so

          \be WRes(D^{2-p})=-\frac{(p-2)c(p)}{12}\int_{X}R\sqrt{g}dx + 
          (p-2)c(p)\int_{X}\frac{3}{2}t_{abc}t^{abc}dx.\ee
This clearly has a unique minimum, given by the vanishing of the torsion term. If we wish to 
          regard the above functional on the affine space of connections, as suggested by Connes, we do 
          the following. Every element of $A_{\s}$ may be written 
          as $(\D_{0}+T)-\D_{0}$, where $\D_{0}$ is 
          the Dirac operator of the Levi-Civita connection. Denote this element by $T$. Then, from what 
          we have proved so far,

          \be
q(T):=WRes(T^{2}\D^{-p})=\frac{(p-2)Vol(S^{p-1})2^{[p/2]}}{p(2\pi)^{p}}\int_{X}\frac{3}{2}t_{abc}t^{abc}
          \sqrt{g}dx.\ee
This is clearly a positive definite quadratic form on $A_{\s}$, for $p>2$, 
 and has unique minimum $T=0$. The 
          value of $WRes(D^{2-p})$ at the minimum is just the other term involving the scalar curvature. 
          Hence, in the even dimensional case, we have completed the proof of the theorem.
       
\subsubsection{The Odd Dimensional Case.}
          For the odd dimensional case ($p=2m+1$) we begin with the observation that 
$\dd^{-p+2}=\D^{-2m}\dd$. As we already know a lot about $\D^{-2}$, the difficult part here will be
the absolute value term. So consider the following 

          \bea\lefteqn{\s_{-p}(|\D|^{2-p})=  \s_{1}(|\D|)\s_{-2m-2}(|\D|^{-2m})  
          + \s_{0}(|\D|)\s_{-2m-1}(|\D|^{-2m})}\nno 
&\qquad &\quad\  \,+\s_{-1}(|\D|)\s_{-2m}(|\D|^{-2m}) 
          -i\sum_{\mu}\partial_{\xi_{\mu}}\s_{1}(|\D|)
          \partial_{x^{\mu}}\s_{-2m-1}(|\D|^{-2m})\nno  
&\qquad &\quad\ \, -i\sum_{\mu}\partial_{\xi_{\mu}}\s_{0}(|\D|) 
          \partial_{x^{\mu}}\s_{-2m}(|\D|^{-2m})\nno
&\qquad &\quad\  \,-\frac{1}{2}\sum_{\mu,\nu} 
          \partial^{2}_{\xi_{\mu}\xi_{\nu}}\s_{1}(|\D|) 
          \partial^{2}_{x^{\mu}x^{\nu}}\s_{-2m}(|\D|^{-2m}).\eea
           This tells us that the only terms to compute are $\s_{1}(\dd),\ 
          \s_{0}(\dd)$ and $\s_{-1}(\dd)$, the other terms having 
          been computed earlier. It is a simple matter to convince 
          oneself that $\s(\dd)$ has terms of integral order only by 
          employing 
          \be \s(\D^{2})=\s(\dd^{2})=\s(\dd)\s(\dd) 
          -i\sum_{\mu}\p_{\xi_{\mu}}\s(\dd)\p_{x^{\mu}}\s(\dd) 
          +\mbox{etc}.\ee
Clearly $\s_{1}(\dd)=\n\xi\n$, which we knew anyway from the 
          multiplicativity of principal symbols. Also

  \bea 
         \lefteqn{ ia^{\mu}\xi_{\mu}+b=2\n\xi\n(\s_{0}(\dd) +\s_{-1}(\dd) 
          +\s_{-2}(\dd))}\nno
&\quad &\quad\   +\s_{0}(\dd)^{2} +2\s_{-1}(\dd)\s_{0}(\dd) \nno 
 &\quad &\quad\  -i\sum_{\mu}\p_{\xi_{\mu}}\s_{1}(\dd)\p_{x^{\mu}}\s_{0}(\dd) 
          -i\sum_{\mu}\p_{\xi_{\mu}}\s_{0}(\dd)\p_{x^{\mu}}\s_{1}(\dd)\nno
&\quad &\quad\  -i\sum_{\mu}\p_{\xi_{\mu}}\s_{1}(\dd)\p_{x^{\mu}}\s_{-1}(\dd) 
          -i\sum_{\mu}\p_{\xi_{\mu}}\s_{0}(\dd)\p_{x^{\mu}}\s_{0}(\dd)\nno
&\quad &\quad\  -i\sum_{\mu}\p_{\xi_{\mu}}\s_{-1}(\dd)\p_{x^{\mu}}\s_{1}(\dd) 
          -\frac{1}{2}\sum_{\mu,\nu}\p^{2}_{\xi_{\mu}\xi_{\nu}}\s_{1}(\dd) 
          \p^{2}_{x^{\mu}x^{\nu}}\s_{1}(\dd)\nno
&\quad &\quad\  -\frac{1}{2}\sum_{\mu,\nu}\p^{2}_{\xi_{\mu}\xi_{\nu}}\s_{1}(\dd) 
          \p^{2}_{x^{\mu}x^{\nu}}\s_{0}(\dd) 
          -\frac{1}{2}\sum_{\mu,\nu}\p^{2}_{\xi_{\mu}\xi_{\nu}}\s_{0}(\dd) 
          \p^{2}_{x^{\mu}x^{\nu}}\s_{1}(\dd)\nno
&\quad &\quad\   +\mbox{order $-2$ or less.}\eea
          Looking at the terms of order $1$, we have 
          \be ia^{\mu}\xi_{\mu}=2\n\xi\n\s_{0}(\dd) 
          -i\sum_{\mu}\p_{\xi_{\mu}}\n\xi\n\p_{x^{\mu}}\n\xi\n,\ee
          or, in Riemann normal coordinates,
          \be \s_{0}(\dd) =_{RN} 
          \frac{1}{2\n\xi\n}ia^{\mu}\xi_{\mu}.\ee
The terms of order $0$ are more difficult, and we find that 
          \bea b=_{RN}2\n\xi\n\s_{-1}(\dd) 
          -\frac{1}{4\n\xi\n^{2}}a^{\mu}\xi_{\mu}a^{\nu}\xi_{\nu}\nno
          -i\xi^{\mu}\p_{x^{\mu}}\s_{0}(\dd) 
          -i\p_{\xi_{\mu}}\s_{0}(\dd)\p_{x^{\mu}}\s_{1}(\dd)\nno
          -\frac{1}{2}\p^{2}_{\xi_{\mu}\xi_{\nu}}\s_{1}(\dd) 
          \p^{2}_{x^{\mu}x^{\nu}}\s_{1}(\dd).\eea
          Remembering that the derivative of an expression in Riemann 
          normal form is not the Riemann normal form of the 
          derivative, we eventually find that 
          \bea b=_{RN,\bmod\n\xi\n} 2\s_{-1}(\dd) 
          -\frac{1}{4}a^{\mu}\xi_{\mu}a^{\nu}\xi_{\nu}\nno 
          +\frac{1}{2}a^{\nu}_{,\mu}\xi_{\nu}\xi^{\mu}
          +\frac{1}{12}\delta^{\mu\nu}R^{\rho\s}_{\mu\nu}\xi_{\rho}\xi_{\s}.\eea          
          In the above, as well as expressing the result in Riemann 
          normal coordinates and $\bmod\n\xi\n$, we have omitted a term proportional to 
          $\xi^{\mu}\xi^{\nu}R^{\rho\s}_{\mu\nu}\xi_{\rho}\xi_{\s}$, 
          since we know that this will vanish when averaged over the 
          cosphere bundle. This gives us an expression for 
          $\s_{-1}(\dd)$ in terms of $a^{\mu}$ and $b$. Indeed, with 
          the same omissions as above we have
          \bea \s_{-1}(\dd)=_{RN,\bmod\n\xi\n}\frac{1}{2}b + 
          \frac{1}{2}a^{\mu}\xi_{\mu}a^{\nu}\xi_{\nu}\nno
          -\frac{1}{8}a^{\nu}_{,\mu}\xi_{\nu}\xi^{\mu} 
          -\frac{1}{24}\delta^{\mu\nu}R^{\rho\s}_{\mu\nu}\xi_{\rho}\xi_{\s}.\eea
          Completing the tedious task of calculation and substitution 
          yields
          \bea \s_{-p}(\dd^{2-p})=_{RN,\bmod\n\xi\n} -\frac{(p-2)}{2}b + 
          \frac{p(p-2)}{4}a^{\nu}_{,\mu}\xi_{\nu}\xi^{\mu}\nno
          -\frac{p(p-2)}{8}a^{\mu}\xi_{\mu}a^{\nu}\xi_{\nu} 
          +\frac{p(p-2)}{24}\delta^{\mu\nu}R^{\rho\s}_{\mu\nu}\xi_{\rho}\xi_{\s}.\eea
We note that the 
          factor $p(p-2)$ arises from $4m^{2}-1=(2m+1)(2m-1)=p(p-2)$.
          \smallskip\newline
          Using the experience gained from the even case, we have no 
          trouble integrating this over the cosphere bundle, giving 
          \be 
          WRes(\dd^{2-p})=-\frac{(p-2)}{12}c(p)\int_{X}R\sqrt{g}d^{p}x 
          +(p-2)c(p)\int_{X}\frac{3}{2}t_{abc}t^{abc}\sqrt{g}d^{p}x.\ee
          Again, this expression clearly has a unique minimum (for $p>1$ and odd) given 
          by the Dirac operator of the Levi-Civita connection.
          \bigskip\newline
          From the results of \cite{KW} and the above calculations, if we twist 
          the Dirac operator by some bundle $W$, the symbol will involve 
          the ``twisting curvature'' of some connection on $W$. This does not influence the 
          Wodzicki residue, and so the above result will still hold, except 
          that the minimum is no longer unique. If we 
          have no real structure $J$, and so are dealing with a 
          spin$^{c}$ manifold, we have the same value at the minimum, 
          though it is now reached on the linear subspace of 
          self-adjoint $U(1)$ gauge terms. This completes the proof of the theorem.
          
 \section{The Abstract Setting}
          \label{pretty}   
          In presenting axioms for noncommutative geometry, Connes has given sufficient 
          conditions for a commutative spectral triple to give rise to a classical geometry, 
but has not given a simple abstract condition to determine whether an algebra has at least one
geometry.
          In our setup we did not try to remedy this situation, but 
          merely to flesh out some of Connes' ideas enough to give 
          the proof of the above theorem. 
          \smallskip\newline
          In light of this proof, we offer a possible characterisation 
          of the algebras that stand a chance of fulfilling 
          the axioms. The main 
          points are that 
          
          1) $\exists c\in Z_{n}(\A,\A\otimes\A^{op})$ such that 
          $\pi(c)=\Gamma$,
          
          2) $\pi(\Omega^{*}(\A))\cong\gamma(Cliff(T^{*}X))$ whilst
          $\pi(\Omega^{*}(\A))/\pi(\delta(\ker\pi))$ is the exterior 
          algebra of $T^{*}X$. This second point may be seen as a consequence of the first order 
condition and the imposition of smoothness on both $\A$ and $\pi(\Omega^*(\A))$. The 
other interesting feature of the (real) representations of $\Omega^*(\A)$, is that if the algebra 
is noncommutative we have self-adjoint real forms, and so gauge terms. With this in mind we should 
regard $\Omega^*(\A)$, or at least its 
representations obeying the first order condition, as a generalised Clifford algebra which includes 
information about the internal (gauge) structure as well. Since this algebra is built on the cotangent space, the 
following definition is natural.
          
          \begin{definition}
          A pregeometry is a dense subalgebra $\A$ of a 
          $C^{*}$-algebra $A$ such that $\Omega^{1}(\A)$ is finite 
          projective over $\A$.
          \end{definition}
This is in part motivated by definitions of smoothness in 
          algebraic geometry, and provides us with our various 
          analytical constraints. Let us explore this.
          \smallskip\newline
          The hypothesis of finite projectiveness tells us that there 
          exist Hermitian structures on $\Omega^{1}(\A)$. Let us 
          choose one, $(\cdot,\cdot)_{\Omega^1}$. We can then extend it to $\Omega^*(\A)$ by 
requiring homogenous terms of different degree to be orthogonal and 
\be (\delta(a)\delta(b),\delta(c)\delta(d))_{\Omega^*}=
(\delta(a),\delta(c))_{\Omega^1}(\delta(b),\delta(d))_{\Omega^1},\ee
and so on. Then we can define a norm on 
          $\Omega^{*}(\A)$ by the following equality:
          \be \n\delta a\n_{\Omega^*}=\n(\delta a,\delta a)\n_{\A}.\ee
          As $(\delta a,\delta a)_{\Omega^1}=(\delta a,\delta a)^{*}_{\Omega^1}$ and 
          $\n\delta a\n_{\Omega^*}=\n(\delta a)^{*}\n_{\Omega^*}$ we have 
          \be \n\delta a(\delta a)^{*}\n_{\Omega^*}=\n\delta a\n^{2}_{\Omega^*}.\ee
          So $\Omega^{*}(\A)$ is a normed $^{*}$-algebra satisfying 
          the $C^{*}$-condition, and so we may take the closure to 
          obtain a $C^{*}$-algebra.
          \smallskip\newline
          What are the representations of $\Omega^{*}(\A)$? Let 
          \be \pi:\Omega^{*}(\A)\rightarrow End(E) \ee
          be a $^{*}$-morphism, and $E$ a finite projective module 
          over $\A$. Thus $\pi|_{\A}$
          realises $E$ as $E\cong \A^{N}e$ for some $N$ and some 
          idempotent $e\in M_{N}(\A)$. 
          \smallskip\newline
  As $E$ is finite projective, we have nondegenerate Hermitian forms and 
          connections. Let $(\cdot,\cdot)_E$ be such a form, and 
          $\nabla_{\pi}$ be a compatible connection. Thus 
          \[ \nabla_{\pi}:E\rightarrow \pi(\Omega^{1}(\A))\otimes E\]
          \[ \nabla_{\pi}(a\xi)=\pi(\delta a)\otimes\xi+ 
          a\nabla_{\pi}\xi\]
          \[ (\cdot,\cdot)_E:E\otimes E\rightarrow \pi(\A)\]
          \be (\nabla_{\pi}\xi,\zeta)_E-(\xi,\nabla_{\pi}\zeta)_E= 
          \pi(\delta(\xi,\zeta)).\ee
          \smallskip\newline
          If we denote by $c$ the obvious map 
          \be c:End(E)\otimes E\rightarrow E,\ee
          then we may define 
          \be \D_{\pi}=c\circ\nabla_{\pi}:E\rightarrow E.\ee
           Comparing 
          \be \D_{\pi}(a\xi)=\pi(\delta a)\xi +aD_{\pi}\xi\ee
          and
          \be \D_{\pi}(a\xi)=[\D_{\pi},a]\xi +aD_{\pi}\xi\ee
          we see that $\pi(\delta a)=[\D_{\pi},\pi(a)]$. Note that 
          $\D_{\pi}$ depends only on $\pi$ and the choice of Hermitian 
          structure on $\Omega^{1}(\A)$. This is because all 
          Hermitian metrics on $E$ are 
          equivalent to 
          \be (\xi,\zeta)=\sum\xi_{i}\zeta_{i}^{*}.\ee
          This in turn tells us that the definition of compatibility 
          with $(\cdot,\cdot)_E$ reduces to compatibility with the above 
          standard structure. The dependence on the structure on $\Omega^{1}(\A)$ arises 
          from the symmetric part of the multiplication rule on 
          $\Omega^{*}(\A)$ being determined by 
          $(\cdot,\cdot)_{\Omega^1}$. If we are thinking of $(\cdot,\cdot)_{\Omega}$ as ``$g$'' 
          in the differential geometry context, then it is clear 
          that $\D_{\pi}$ should depend on it if it is to play the 
          role of Dirac operator. Thus it is appropriate to define a 
          representation of $\Omega^{*}(\A)$ as follows.
           
          \begin{definition}
          Let $\A\subset A$ be a pregeometry. Then a representation 
          of $\Omega^{*}(\A)$ is 
          a $^{*}$-morphism $\pi:\Omega^{*}(\A)\rightarrow End(E)$, 
          where $E$ is finite projective over $\A$ and such that the first order condition holds. 
          \end{definition}     
In the absence of an operator $\D$, we interpret the first order condition as saying that
$\pi(\Omega^0(\A))$ lies in the centre of $\pi(\Omega^*(\A))$, at least in the commutative case. In
general, we simply take it to mean that the action of $\pi(\A^{op})$ commutes with the action of
$\pi(\Omega^*(\A))$.      
Next, it is worthwhile pointing out that representations of $\Omega^*(\A)$ are a 
good place to make contact with Connes description of cyclic cohomology via cycles, \cite{C}, 
though this will have to await another occasion. In this definition we encode the first order condition 
by demanding that 
 $\pi(\Omega^*(\A))$ is a symmetric $\pi(\A)$ module in the commutative case. In the 
noncommutative case that we discuss below, we will require that the image of $\A^{op}$ 
commutes with the image of $\Omega^*(\A)$.  Let us 
           consider the problem of encoding Connes' axioms in this 
           setting.
           \smallskip\newline
           The first thing we require is an extension of these 
           results to $\Omega^{*}(\A)\otimes\A^{op}$. Since a left 
           module for $\A$ is a right module for $\A^{op}$, we shall 
           have no problem in extending these definitions if we 
           demand that $[\pi(a),\pi(b^{op})]=0$ for all $a,b\in\A$. 
           Since $\overline{\Omega^{*}(\A)}$ is a $C^{*}$-algebra, any 
           representation of it on Hilbert space lies in the bounded 
           operators. This deals with the first two items of 
           Definition 2 in section 3. The real structure will clearly remain 
           as an independent assumption. What remains? 
           \smallskip\newline
           We do not know that $\A$ is ``$C^{\infty}(X)$'' in the 
           commutative case yet. Examining the foregoing proof, we 
           see that we first needed to know that the elements involved in 
           the Hochschild cycle $c$ generated $\pi(\A)$, which came from 
           $\pi(c)=\Gamma$. Then we needed to show that the condition 
           \be \pi(a),\ 
           [\D_{\pi},\pi(a)]\in\cap^{\infty}Dom(\delta^{m})\ee
           implied that $\pi(a)$ was a $C^{\infty}$ function and that $\pi(\Omega^*(\A))$ was the 
smooth sections of the Clifford bundle. Recall that 
           $\delta(x)=[|\D_{\pi}|,x].$
           \smallskip\newline
So having a representation, we obtain $\D_{\pi}$, and we 
           can construct $\dd_{\pi}$ if $\D_{\pi}$ is self-adjoint. 
           This will follow from a short computation using the fact that $\nabla_{\pi}$ is 
           compatible.
           \smallskip\newline
           Then we say that $\pi$ is a smooth representation if 
           \be 
           \pi(\Omega^{*}(\A))\subset\cap^{\infty}Dom(\delta^{m}).\ee
This requires only the finite projectiveness of $\Omega^*(\A)$ to state, though this is not
necessarily sufficient for it to hold.
           As $E\cong e\A^{N}$, this also ensures that 
           $\D_{\pi}:E\rightarrow E$ is well-defined. Further, in the 
           commutative case we see immediately that $\D_{\pi}$ is an 
           operator of order $1$. Thus any pseudodifferential 
           parametrix for $|\D_{\pi}|$ is an operator of order $-1$. We 
           can then use Connes' trace theorem to state that 
           $|\D_{\pi}|^{-p}\in\LL^{(1,\infty)}$. The imposition of 
           Poincar\'{e} duality then says that $\intcross|\D_{\pi}|^{-p}\neq 0$. 
           \smallskip\newline
           So, a pregeometry is a choice of ``$C^{1}$'' functions on a 
           space. Given a first order representation $\pi$ of the universal 
           differential algebra of $\A$ provides an operator 
           $\D_{\pi}$ of order $1$. We use this to impose a further 
           restriction (smoothness) on the representation $\pi$ and algebra $\A$. 
           
 \begin{definition}
           Let $\pi:\Omega^{*}(\A)\rightarrow End(E)$ be a smooth 
           representation of the pregeometry $\A\subset A$. Then we 
           say that $(\A,\D_{\pi},c)$ is a $(p,\infty)$-summable 
           spectral triple if 
           
           1) $c\in Z_{p}(\A,\A\otimes\A^{op})$ is a Hochschild cycle with $\pi(c)=\Gamma$
           
           2) Poincar\'{e} duality is satisfied
           
           3) $E$ is a pre-Hilbert space with respect to 
           $\intcross (\cdot,\cdot)_E\dd_{\pi}^{-p}$.
           \end{definition}
           
           \begin{definition}
           A real $(p,\infty)$-summable spectral triple is a 
           $(p,\infty)$-summable spectral triple with a real structure.
           \end{definition}
           It is clear that this reformulation loses no information.
\smallskip\newline           
This approach may be helpful in relation to the work of \cite{FGR}. By employing extra operators and 
imposing supersymmetry relations between them, the authors show that all classical forms of differential 
geometry (K\"{a}hler, hyperk\"{a}hler, Riemannian...) can also be put into the spectral format. Examining their 
results show that the converse(s) may also be proved in a similar way to this paper, provided the correct axioms
are provided. The elaboration of these axioms may well be aided by the above formulation, but this will have to
await another occasion.
\section{Acknowledgements}
I would like to take this opportunity to thank Alan Carey for his support and assistance whilst writing this paper,
as well as  Steven Lord and David Adams for helpful discussions.  I
am also indebted to A. Connes and J.C. Varilly for pointing out serious errors and omissions in the original version of
this paper, as well as providing guidance in dealing with those problems.

\end{document}